\documentclass[10pt,conference]{IEEEtran}
\usepackage{version}
\excludeversion{nsdi}
\excludeversion{techreport}
\includeversion{security}
\usepackage[latin1]{inputenc} %
\usepackage{cite} 
\usepackage{amsfonts}
\usepackage{xspace}
\usepackage{graphicx}
\usepackage{subfigure}
\usepackage{multirow}
\usepackage{amsmath}
\usepackage{color} %
\usepackage{epsfig}
\usepackage{fancybox}
\usepackage{calc}
\usepackage{url}
\usepackage{balance}
\usepackage{times}

\newtheorem{theorem}{Hypothesis}

{\begin{list}{}%
         {\setlength{\leftmargin}{#1}}%
         \item[]%
}
{\end{list}}

\newcommand{\paragraphb}[1]{\vspace{0.03in}\noindent{\bf #1} }
\newcommand{\paragraphe}[1]{\vspace{0.03in}\noindent{\em #1} }

\newcommand{\prateeknew}{\textcolor{black}}

\newcommand{\prateek}{\textcolor{black}}
\newcommand{\prateekccs}{\textcolor{black}}
\newcommand{\dongho}{\textcolor{black}}
\newcommand{\matt}[1]{{\color{black}{#1}}}

\newcommand{\cut}[1]{}
\date{~}
\title{\Large \bf Mirage: Towards Deployable DDoS Defense for Web Applications}
\author{
    \IEEEauthorblockN{Prateek Mittal}
    \IEEEauthorblockA{University of California, Berkeley \\ pmittal@eecs.berkeley.edu}
    \and
    \IEEEauthorblockN{Dongho Kim}
    \IEEEauthorblockA{Cisco Systems \\ dkim99@illinois.edu}
    \and
    \IEEEauthorblockN{Yih-Chun Hu, Matthew Caesar}
    \IEEEauthorblockA{University of Illinois at Urbana-Champaign \\ \{yihchun,caesar\}@illinois.edu}
}

\begin{document}

\maketitle
\begin{abstract}

\prateeknew{Distributed Denial of Service (DDoS) attacks form a serious
threat to the security of Internet services.}  However, despite over a decade of
research, and the existence of several proposals to address this problem, there
has been little progress to date on actual adoption.  
\prateek{We present Mirage, 
a protocol that achieves comparable performance to other DDoS
mitigation schemes while \matt{providing 
benefits} when deployed only in the server's local 
network and its upstream ISP, where local business objectives may incentivize deployment.}
Mirage does not require source end hosts to install any software to access 
Mirage protected websites.
\begin{techreport}
Unlike previous proposals, Mirage only requires functionality from routers that is already 
deployed in today's routers, though this functionality may need to be scaled 
depending on the point of deployment.
\end{techreport}


Our approach 
is that end hosts can thwart the
attackers by employing the principle of a moving target: end hosts in our
architecture periodically change IP addresses to keep the attackers guessing.
Knowledge of an active IP address of the destination end host can act as an
implicit authorization to send data.
\prateek{ We evaluate Mirage using theoretical analysis, 
simulations and a prototype implementation on PlanetLab.  
We find that our design provides a first step towards a deployable, yet effective 
DDoS defense.}

\end{abstract}

\section{Introduction}
\label{sec:intro}

\prateeknew{Denial of service (DoS) attacks form a serious threat to 
the security of Internet services.} In a DoS attack, a single computer or a group of
computers (DDoS) flood the victim's machine by sending a large number of
packets, exhausting its bandwidth capacity (\prateek{or software processing capabilities}).
This causes legitimate traffic to back off,
resulting in denial of service to properly behaving hosts.
\prateeknew{The frailty of today's Internet to DDoS attacks is evident by
recent attacks on entire countries such as Georgia and Estonia,
whistle-blowing websites such as Wikileaks, financial transaction hubs 
such as Mastercard, Visa, and PayPal, intelligence agencies such as CIA, 
government organizations such as US department of justice, and online 
election servers~\cite{donner:sp07,wikileaks-ddos,visa-ddos}. }

There are two main schools of thought regarding DDoS defense. \emph{Filtering}-based 
approaches~\cite{liu:sigcomm08,argyraki:usenixatc05,mahajan:ccr02} first
aim to detect and classify the malicious traffic and then install filters at
the network layer to prevent the malicious traffic from exhausting victim's
resources. On the other hand, \emph{Capability}-based
approaches~\cite{yang:sigcomm05,yaar:sp04,parno:sigcomm07} aim to
send only authorized traffic to hosts in the Internet.
\dongho{%
Extensive academic work has been done on these %
approaches, and
their tradeoffs are well understood~\cite{liu:sigcomm08}. However,
despite over a decade of research, little progress has been made on
real-world deployment. \prateeknew{In fact, the number of DDoS 
attacks increased by 57\% in 2011~\cite{kaspersky}.} A fundamental 
obstacle to deployment is that
a deploying AS does not benefit until other ASes %
\prateek{also participate in the}
scheme. This dependency between ASes is a chicken-and-egg problem
for deployment in the Internet, which traditionally %
does not force a protocol to be deployed.
\prateek{%
Phalanx~\cite{dixon:nsdi08} %
aims to reduce the number of ASes that are required to deploy the
protocol for effective %
defense, but does not completely solve
the chicken-and-egg problem (see \S~\ref{sec:related}).}}
\prateek{ In this paper, we propose {\em Mirage}, a DDoS defense mechanism that aims to %
remove the
required deployment at other ASes as much as possible. Towards this
end, %
we adopt an approach similar to %
{\em frequency hopping} in
wireless networks~\cite{freqhop}: in Mirage, a destination end host %
varies its IP addresses according to a
pseudorandom sequence known only to authorized hosts.}
Knowledge of an {\em
active} IP address owned by the destination end host acts as an
implicit authorization to send data, and  \dongho{enables Mirage to
reduce the required deployment of other ASes.
To realize this idea in a concrete system, we %
leverage known ideas: computational puzzle~\cite{wang:sp03,wang:ccs04,halderman:ccs07}, filtering and fair queueing.
\prateek{Our key contribution is a system architecture that integrates
these existing primitives with the novel paradigm of IP address hopping,
with the goal of reducing the need for deployment across organizational
boundaries.}
The security of Mirage depends on the space of IP addresses that a
destination end host can use. To get a sufficiently large space of
IP addresses, each Mirage-protected server uses its IPv6 prefix and
chooses interface addresses as these secret addresses. The
practicality of Mirage will grow each day with the increasing
deployment of IPv6 and the ability of IPv4-only-connected hosts to
use IPv6 through tunneling, as described in \S~\ref{sec:limit}.
Randomly hopping IP addresses play the same role as capabilities
in capability-based schemes, which in contrast to Mirage, introduce
a completely new packet header~\cite{yaar:sp04,yang:sigcomm05,parno:sigcomm07}.}
\prateekccs{It is much more likely that IPv6
will witness widespread deployment, as opposed to a new 
custom Internet header.} 

\prateek{ We evaluate Mirage using theoretical analysis, large-scale
simulations and a prototype implementation on PlanetLab. We find
that Mirage is able to provide honest nodes with their fair share of
system resources, comparable to previous DDoS mitigation mechanisms.
At the same time,  Mirage is able to provide some benefits to a
server when only the server's local network and its upstream ISP
deploy it. Unlike previous proposals, Mirage only requires
functionality from routers that is already deployed in today's
routers, though this functionality may need to be scaled depending
on the point of deployment.
In particular, Mirage does not %
require \matt{router hardware or software modifications} to support
new headers/fields in network packets, and achieves this without
changing the semantics of existing header fields. Our architecture
also does not require any cryptography at the routers. Finally,
source end hosts can take advantage of Mirage-protected sites
without installing any software. We find that our design provides a
first step towards a deployable, yet effective DDoS defense. }

\begin{techreport}
We start by discussing related work (\S~\ref{sec:related}) and
our overall design (\S~\ref{sec:overview}). We then
present Mirage in detail (\S~\ref{sec:protocol} and~\ref{sec:deployment}),
discuss its security properties (\S~\ref{sec:security}), and evaluate
our design with simulations and a prototype implementation (\S~\ref{sec:evaluation}).
Finally, we present a comparison of Mirage's deployment properties with
previous proposals, and conclude with a discussion of ramifications (\S~\ref{sec:discussion}
and~\ref{sec:conclusion}).
\end{techreport}

\section{Related Work}
\label{sec:related}

Our work is related to two main schools of thought about DDoS mitigation:
{\em filtering} and {\em capability}-based designs.

\paragraphb{Filtering and rate limiting:}
Pushback~\cite{mahajan:ccr02} defends against
DDoS attacks by having congested routers rate limit the set of flows
responsible for congestion (aggregate rate limiting). If local rate
limiting is not sufficient to mitigate congestion/DDoS attacks, then
the router contacts its upstream router and asks it to perform rate
limiting as well. %
AITF~\cite{argyraki:usenixatc05} is a traffic filtering architecture
that leverages record routing techniques to enable a victim to
identify routers close to a source. The victim can then install
filters close to the source of attack traffic.
StopIt~\cite{liu:sigcomm08} proposes a filtering mechanism whereby a
receiver can request a particular flow to be blocked for a period of
time. %
While filtering-based schemes are powerful, they face significant practical
hurdles, as they assume the ability to
differentiate attack traffic from legitimate traffic.
Adversaries may make use of cover traffic
to hide their communication, or may mimic legitimate traffic
(e.g., by mimicking legitimate GET requests).
\prateek{Mirage \matt{does} not make any assumptions about the ability
to classify malicious traffic;
however, if %
a method to differentiate attack traffic is available, Mirage can leverage it %
to provide a significant subset of the requisite
filtering capability.}

\paragraphb{Capabilities:} Capability-based schemes provide the receiver
an ability to directly control its reachability within the network.
When a sender wants to transmit packets to a receiver, the
receiver decides whether to permit it and lets routers know its
decision. Then, the router allows the corresponding packets to pass.
SIFF~\cite{yaar:sp04} gives a technique to efficiently allow receivers
to authorize senders to transmit.
A router sends tokens to a receiver that can be used to authorize sender
requests. When a receiver wants to receive packets from the sender,
the receiver passes the token to a sender. The sender adds the
token to packet headers it transmits to get a preferential service from the router.
TVA~\cite{yang:sigcomm05} refines SIFF to defend against brute-force
attacks to improve practicality and efficiency.
Portcullis~\cite{parno:sigcomm07} addresses a vulnerability in
both SIFF and TVA where an attacker could launch a denial of
capability attack, and prevent the initial request packets of honest users
from reaching the victim. Portcullis makes use of computational
puzzles to force requesting users to perform work before being able
to access the initial request channel. %
In \S~\ref{sec:deployment-comparison}, we discuss the deployment
challenges for capability-based schemes, including the problem of
low benefit for early adopters,
\matt{the need to upgrade client end-host software},
\matt{requiring new router primitives such as cryptographic support},
\prateekccs{as well as requiring a new custom Internet header}.
Additionally,
we note that in most of the above approaches, the capability to send a
packet is specific to the path used to route a packet to a destination.
Thus, route changes in the Internet pose a significant challenge for
these approaches. In contrast, Mirage uses the knowledge
of an active destination IP address within the IP address range of
the destination end host as an implicit authorization to send data.
Such an authorization mechanism is end-to-end, and is not affected
by route changes in the Internet.

\paragraphb{Other approaches:} %
Phalanx~\cite{dixon:nsdi08} proposes to use a {\emph swarm}, a large
pool of geographically and topologically distributed well
provisioned machines to absorb the DDoS attack, and forward
legitimate traffic to the victim. Once a few ASes deploy an adequate 
number of such well provisioned machines to form an overlay network, any ISP can 
gain immediate benefit from deploying Phalanx. While the design of Phalanx 
is a significant improvement over prior work in this domain, it still 
requires other ASes to deploy the protocol and participate in an overlay network. 
Wang et al.~\cite{wang:sp03,wang:ccs04} have also proposed the use
of client puzzles, but their approach does not focus on deployment.
NetFence~\cite{liu:sigcomm10} proposes to
use the network as the first line of defense to mitigate DDoS attacks, 
but assumes that IP spoofing is not possible. 
Speak-up~\cite{walfish:sigcomm06} uses proof of work techniques for
DDoS defense, but its focus is on application-layer attacks. \prateeknew{Similarly, 
the end-to-end approach of Gligor~\cite{gligor:03} is also limited to 
application-layer 
attacks.} Kandula
et al.~\cite{kandula:nsdi05}, and Morein et al.~\cite{morein:ccs03} 
propose defense mechanisms wherein users solve CAPTCHAs~\cite{captchas} to differentiate themselves 
from bots. Mirage does not require users to
solve CAPTCHAs. %
Keromytis et al.~\cite{keromytis:sigcomm02} propose an approach
where only select nodes in an overlay network are allowed to
communicate with a destination, and their IP addresses are kept
hidden. Pre-authorized users are given knowledge of a subset of
those select nodes, and can use them as a proxy to communicate with
a destination. 
Mirage builds upon the idea of hiding proxy IP
addresses by introducing the notion of destination IP address
hopping. Mirage does not require the use of proxies or overlay
networks, and provides service to all users (not just pre-authorized
users). \prateekccs{Similar to Mirage, Stavrou et al.~\cite{stavrou:ccs05} also 
propose the use of spread-spectrum techniques, 
but they rely on an external overlay network, and do 
not focus on the issue of deployment.}

\paragraphb{Moving target defenses:} \prateeknew{Mirage is inspired from frequency 
hopping in wireless networks, and more generally, it is related to 
the concept of moving target defenses~\cite{jajodia-moving-target11} which 
aim to create uncertainty for the adversary. Mirage's notion of IP address 
hopping is related to the concept of address space randomization techniques 
such as ALSR~\cite{shacham:ccs04}, which aims to defend against untrusted code, 
and NASR~\cite{nasr}, which aims to defend against hit-list worms, and the work 
of Shue et al.~\cite{shue:ccr12}, which aims to implement network capabilities. 
The key challenges in the application of address space randomization techniques 
for DDoS defense are to (a) enable (any) legitimate client to access the server, 
(b) constrain the effects of malicious clients, and (c) defend against attacks 
on upstream network prefixes. Mirage solves these challenges, with the additional 
goal of reducing the need for deployment across organizational boundaries. 
In contrast, both NASR and the work of Shue et al. are vulnerable to malicious 
clients that query DNS servers to learn victim server's address; unlike Mirage, 
they lack proof-of-work mechanisms such as computational puzzles to limit attackers. 
Moreover, both honest and malicious clients are issued identical addresses for the 
victim server, enabling malicious clients to dominate victim's network resources. 
Finally, both NASR and Shue et al. do not defend against DDoS attacks on 
victim's upstream network prefixes, for example, an adversary can flood an 
IP address that was previously in use by the victim. 
}

\begin{techreport}
\subsection{Mirage Properties}
\label{sec:properties}

Mirage takes an end-to-end approach, which brings several benefits.
First, Mirage does not require software or hardware changes to routers.
The functionality Mirage requires is to a large extent already deployed
in existing networks, and can be realized with configuration changes
to existing infrastructure.
Mirage does not place additional complexity or computational burdens on routers (such as cryptography), and works correctly in the presence of route changes.
Source end hosts can take advantage of Mirage protected sites without installing any software.
Mirage also does not depend on the deployment of mechanisms that have proven hard to deploy in the network,
such as source address spoofing defense%
, and provides benefits
even in the presence of compromised routers.
\end{techreport}

\section{Mirage Overview}
\label{sec:overview}

We next describe some goals and limitations of our work (\S~\ref{sec:limit})%
\prateekccs{, and explain how we design Mirage to achieve these goals (\S~\ref{sec:arch}).}

\subsection{Design Goals and Limitations}
\label{sec:limit}

\begin{techreport}
The design goals of our architecture are as follows.
\end{techreport}

\prateek{
\paragraphe{1. Incremental and incentivized deployment:} We target an incrementally
deployable design where a small number of routers should be able to deploy Mirage
and bring immediate benefit to downstream servers. We note that in such a design,
local business objectives may incentivize deployment. Our architecture
should not interfere with operation of existing or non-upgraded network protocols or
systems. %
Mechanisms that introduce additional packet headers or protocol layers
may not satisfy this property. For instance, schemes that introduce additional headers can face
significant challenges during incremental deployment, due to incompatibilities \prateek{with}
scrubbing and IDS services, layer-7 and layer-4 load balancing,
and other middleboxes and network services that inspect non-IP layers of the network stack.
}

\prateek{
\paragraphe{2. Lowering deployment cost:} We target a design which minimizes the requirement for
cooperation across administrative or trust boundaries, and does not rely on an external overlay
network to deploy the mechanism. Mechanisms that require the use of trusted hardware~\cite{anderson:sigcomm08},
or require end users to upgrade software face major deployment hurdles and %
are incompatible with our design goals. Similarly,
we avoid reliance on new primitives from routers, such as router cryptography and additional packet headers,
which require significant support from vendors and increase the cost of network equipment.
}

\paragraphe{3. Network fairness:} In the absence of techniques to classify attack traffic,
we aim to provide each user with its fair share of the network. However, if techniques to partially classify attack traffic
are available, then Mirage can leverage them. %
\paragraphe{4. Low overhead:}
We target a design that does not impose additional cost on the network
or end hosts when the system is not under attack (adaptive defense).

\paragraphb{Limitations:} Our work has several limitations.
First, like other schemes that use puzzles, such as Portcullis~\cite{parno:sigcomm07} and Phalanx~\cite{dixon:nsdi08},
Mirage requires the ability to distribute cryptographic information (puzzles)
in a manner that is not subject to DDoS. Mirage can use existing replicated
services such as Akamai or well provisioned cloud services such as Amazon S3 or
even DNS.
Second, Mirage requires a large IP address
space to perform effectively. To achieve this, Mirage can make use of IPv6.
The Internet is already in a transition phase to IPv6~\cite{world-ipv6-day}, and we note that
IANA has already exhausted its pool of IPv4 allocation blocks in February 2011~\cite{iana}. %
Mirage is able to accommodate scenarios
where source end hosts are in an IPv4 only network, by the use IPv4-IPv6 translators.
For example, Teredo IPv6 tunneling~\cite{teredo} is already built into Windows.
Mirage can even protect destination end-hosts serviced by an IPv4 network;
in such scenarios, the victim sets up a tunnel to a IPv6 provider that supports Mirage.
To protect against native IPv4 DDoS, the victim either keeps its IPv4 address secret,
or requests its ISP to block any traffic that does not originate from the tunnel server.
We emphasize that in contrast to Mirage's approach of leveraging IPv6, prior work
on capability advocates a completely new packet header. 

Third, our design requires loose time synchronization, on the order
of tens of seconds. To address this, Mirage may require
external systems such as NTP, which can provide time synchronization
with accuracies on the order of hundreds of milliseconds on wide
area networks~\cite{ntpstudy}. Fourth, Mirage targets per
computation fairness. If the ratio of honest sources' computational
power to the total computational power of all nodes (honest nodes
and attacker nodes) is $r$, then Mirage provides the honest nodes
with a fraction $r$ of the bottleneck link bandwidth. In scenarios
where there is a mismatch between the computational resources of the
attacker and the honest nodes (for example, a webserver with a small
user base being attacked by a very large botnet), the fair share of
the honest nodes will be small.\footnote{\prateekccs{Even though the fair share of 
honest nodes may be small in some scenarios, their requests will 
still succeed.}} We note that this limitation is also
shared by other state of the art mechanisms that rely on
computational puzzles, such as Portcullis~\cite{parno:sigcomm07} and
Phalanx~\cite{dixon:nsdi08}. Another aspect related to the issue of
resource mismatch is the differences in computational capabilities
of various devices, such as a smartphone and a %
GPU.
\matt{
However, 
Mirage is compatible with the use of memory-bound puzzles~\cite{memory-bound-puzzles}, which have been shown to lower resource disparity between devices.
}

\paragraphb{Threat model:}
\prateekccs{There are two types of DoS attacks: (a) network-layer attacks, 
where an attacker attempts to overwhelm the transmission capabilities of 
the underlying network, and
(b) application-layer attacks, where the attacker attempts to overwhelm
processing capabilities of the victim's application software.
While the focus of this paper is on network-layer DoS attacks, 
Mirage provides an interface to the application layer to fair queue 
incoming requests based on proof of work (destination IP addresses).} 
\prateeknew{Similar to Portcullis~\cite{parno:sigcomm07} and Phalanx~\cite{dixon:nsdi08}, 
we assume that lookup services such as DNS can be highly replicated (since 
they serve short, static content), and are not subject to DDoS.} 
We assume the adversary may have access to a large number of hosts
(e.g., via ownership of a botnet), and may perform IP spoofing.
We assume routers can be compromised, though we note that a compromised
router can always block traffic towards the victim.

\subsection{System architecture}
\label{sec:arch}

\begin{figure}
\includegraphics[width=3in]{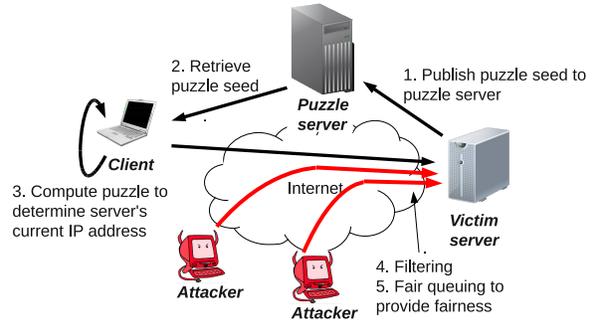}
\begin{security}
\vspace{-0.1in}
\end{security}
\caption{Mirage network architecture}
\label{fig:arch}
\begin{security}
\vspace{-0.2in}
\end{security}
\end{figure}

\paragraphe{IP address hopping:} To mitigate denial of service attacks, we need some way to make it harder for
the attacker to reach the server.
 \dongho{In previous approaches, capabilities or randomly chosen proxies (Phalanx~\cite{dixon:nsdi08}) are used for that purpose.
 However, these approaches add a new packet header and require deployment in other ASes thereby hindering incentivized deployment.
Mirage's approach is to use randomly changing destination IP addresses. This idea enables Mirage to avoid the necessity of other ASes' deployment.}
Mirage consists of a set of add-ons to existing Internet services (Figure~\ref{fig:arch}) that enable a server
to dynamically {\em hop} (change) its IP address.
 In particular, the server \matt{is assigned a {\em set} of IP addresses, and} repeatedly
modulates
\matt{the address it uses from this set} via a deterministic pseudorandom function.  This function
computes the server's current IP address given the  current time as input.  The
server can then share that pseudorandom function with authenticated clients;
these clients can then determine the %
IP address used by the server by
computing the pseudorandom function.

\paragraphe{Slowing the attacker with computation-limited hopping functions:}
While certain Internet services can authenticate clients (e.g.,
systems that use CAPTCHAS or require the user to sign in with a
secret password), other services may be unable to distinguish valid
user requests from malicious requests.  To support these services,
\dongho{a form of fair sharing mechanism is necessary. Options for
fair sharing include bandwidth-based
fairness~\cite{walfish:sigcomm06} or computation-based
fairness~\cite{parno:sigcomm07}. We do not use bandwidth-based
fairness since it may induce congestion collapse in network.} Mirage
makes it more computationally
difficult for the client to retrieve the %
IP address using the pseudorandom function.  This is done by having
the server construct a computationally-harder version of the
pseudorandom function, which is then handed out to clients.  This
computationally-harder version is constructed by incorporating a
cryptographic puzzle into the pseudorandom function returned to the
client. Mirage makes use of hard-to-DoS infrastructures such as
Akamai, DNS \prateek{or well provisioned cloud services such as
Amazon S3} for puzzle distribution. \dongho{It is important to note
that putting all servers to cloud services to defend against DDoS
attacks is expensive. Mirage uses such service for only puzzle
distribution.}

\paragraphe{Avoiding network bottlenecks with in-network filtering:}
Even though a server can perform IP address hopping, an adversary can still
\matt{attack the server by sending traffic to any of the alternate IP addresses assigned to
the server, even though they are not in use.
To address this, Mirage  %
requires routers %
to perform in-network filtering.} In particular, the victim host
may instruct upstream routers of IP address ranges that it is currently not using.
This may be done by a variety of mechanisms. %
For simplicity, our design assumes that the end host explicitly publishes access
control lists that are serviced by upstream routers.
\matt{
Ideally, this filtering should be done near bottleneck links.
}

\paragraphe{Isolating the attacker with \matt{address sets}:} \prateek{Hopping IP addresses}
to evade the attacker only works when the attacker does not know
the new IP address of the server.  
\prateekccs{However, preventing the attacker from knowing an active IP address of 
the victim seems difficult, while still enabling non-malicious clients to 
know the current IP address.}
Hence, Mirage instead attempts to constrain the effects of malicious clients by 
associating a {\em set} of active IP addresses with the server, and and returning 
different elements of that set to different clients.  
To achieve this,  we leverage standard techniques
(such as those used by DNS~\cite{akamai}) to return different IP addresses to 
different clients based on their topological location.
\prateek{To defend against attacks where a single malicious client
solves a computational puzzle and shares the puzzle solution with its
colluding attackers, we need to provide isolation across elements of
the set. To do this, the server or the bottleneck link can utilize fair queuing. %
}
\begin{techreport}
\prateek{We note that most current mechanisms attempt to isolate the attacker
based on load balancing on \emph{source} IP addresses. However, given that IP spoofing is
possible in today's networks, the attackers can easily game such mechanisms. Mirage's
novelty lies in enabling a load balancing mechanism based on \emph{destination}
IP addresses.}
\end{techreport}

An example of the Mirage protocol is shown in Figure~\ref{fig:arch}: (1) First,
the server determines a seed for a globally-known pseudorandom
function, and registers it with its puzzle server.  (2)
The server then begins using that pseudorandom function to compute
its currently active set of IP addresses. This function is
periodically recomputed to perform the hopping.
(3) When a host wishes to make a request, it does a DNS
lookup for the victim, gets redirected to the puzzle server,
and retrieves the computational puzzle. The user then executes
the puzzle to determine the server's current IP address. The user
then sends packets to the server. The user periodically re-executes
the puzzle to keep track of the server's current IP address.  (4)
When a malicious host, such as a bot, wishes to
execute a request, it follows exactly the same procedure. Due to the
puzzles, the botnet cannot acquire more IP addresses than its
computational power.
\prateek{Finally, the network filters traffic that does not have a victim
server's current IP address.
}
Due to the per-destination fair
queuing, bandwidth is shared fairly across the destination IP addresses,
reducing power of an attacker.

\section{Protocol Description}
\label{sec:protocol}

In this section, we describe the basic Mirage design. In particular, we describe the
higher level functions performed by destination %
end hosts, %
network routers, %
puzzle servers, %
and source end hosts. %
\prateeknew{We discuss security/performance improvements, as well as deployment details of our protocol in
\S~\ref{sec:deployment}}.

\subsection{Destination end hosts}
\label{sec:destination}

\paragraphb{IP address hopping:} Our core mechanism is inspired by
{\em frequency hopping}, used in wireless networks. Frequency hopping
is a way to transmit radio signals by rapidly switching the carrier
through a pseudorandom sequence of channels. If a receiver knows
the pseudorandom sequence (e.g., if it shares a key with the sender),
it can listen on the same sequence of frequencies to receive the transmission.
Doing this can make wireless devices less susceptible to ``jamming'' attacks,
which attempt to deny service by transmitting undesired traffic on a communication
channel.

In our work, we apply this idea to make end hosts less susceptible to
denial of service attacks. In particular, end hosts
are assigned a consecutive range (prefix) of IP addresses.
The sender and receiver then periodically hop through a pseudorandom sequence of IP addresses
when communicating.
We assume IPv6, to improve feasibility of allocating multiple IP addresses to end hosts, given
IPv6's larger address space (IPv6 addresses are 128 bits, with 64 bit subnet/interface addresses). 
\footnote{\prateeknew{Mirage permits the same number of IPv6 devices as in the current IPv6 deployment model, 
where devices are assigned a unique 64-bit interface address (RFCs 3513 and 4291~\cite{rfc3513,rfc4291}).}}
More generally,
instead of having a single active IP address, end hosts can choose to have a
\emph{set} of active IP addresses from amongst their allocated range. This set
of active IP addresses will periodically change. Each end host performs IP
address hopping using a local cryptographic master key. At any instance of
time, the master key can be used to determine the set of active IP addresses
for that end host: %

\[ IP(i) = PREFIX || H( ENC_{KEY}(i || TIME)) \]

\noindent where $IP(i)$ denotes the end host's $i$'th active IP address, $PREFIX$ is the prefix IP address associated with the
end host, $||$ represents the concatenation operation, $H$ is a cryptographically secure hash function with output length $128-|PREFIX|$ bits, and $ENC_{KEY}(x)$
is the  encryption of $x$ with the key $KEY$.
This set of active IP addresses is kept a secret, and will
only be used under a DDoS attack. In addition to this secret set of IP
addresses (which keep changing), the end hosts also maintain a static (not
hopped) IP address, which is used when the host is not under %
attack.

\paragraphb{Puzzle server redirection:} As in the current Internet architecture, destination end hosts
set up a DNS entry for their hostname. Under a DDoS attack, a destination end host (victim server) re-registers
with its authoritative DNS (ADNS) to point its record to the puzzle server, using
similar techniques to those used in CDNs for redirection and load balancing (e.g., small TTL).
From that point onwards, the source end hosts are
redirected to the puzzle servers.

\paragraphb{Filtering requests to routers:} Under a DDoS attack, the victim
communicates the current set of active IP addresses to its upstream \prateek{ASes}.
The upstream \prateek{ASes} can then block incoming traffic to any of the remaining
IP addresses which are not in the active set. Thus, knowledge of any active IP address
can be used as an implicit authorization to send data to that destination end host.

\paragraphb{Leveraging attack traffic identification:}
\prateek{
Note that in the absence
of the ability to identify attack traffic, Mirage aims to provide per computational
fairness to clients. However, if it is possible to identify attack traffic,
then Mirage can leverage this ability to improve performance for honest clients.
In particular, the victim server can stop using the destination IP addresses corresponding to
attack traffic, and request the upstream ASes to filter those IP addresses.
}

\subsection{Routers}
\label{sec:routers}

\paragraphb{\matt{Filtering traffic with existing router interfaces}:}
Whenever a victim is under a DDoS attack, the
upstream routers receive information about the set of active IP addresses
currently in use by the victim. The upstream routers can then
filter out incoming traffic to all remaining (non-active) IP
addresses \prateek{(as well as other IP addresses that are identified
by the victim as being associated with attack traffic)}.
Note that the victim can start by installing filters at its
edge router, and continue this process at the routers further upstream
until its link is no longer congested.  %
\prateek{In \S~\ref{sec:deployment}  we show that the upstream \prateek{ASes} can
perform such filtering using \emph{existing} router interfaces \matt{such as
access control lists (ACLs) configured via iBGP feeds from IRSCP~\cite{irscp} style } management servers.
Moreover, in \S~\ref{sec:evaluation}, we show that performing such filtering
only at the server's upstream provider is sufficient to block most attacks.
}

\paragraphb{Per-destination fair queueing:} After filtering out packets with
non-active destination IP addresses, the upstream routers perform fair queueing per
destination. This results in per destination address fair allocation of resources
when attack traffic cannot be distinguished from legitimate traffic. We note that
routers already support fair queueing as an option, easily enabled via configuration
changes. We discuss this in more detail in \S~\ref{sec:deployment}.

\begin{techreport}
The combination of these techniques results in the attack traffic being
filtered before the bottleneck link, as well as fair allocation of bandwidth
resources.
\end{techreport}

\subsection{Puzzle Servers}
\label{sec:puzzleservers}
Under attack, source end hosts are redirected to a puzzle server.

\paragraphb{Active IP address generation:}
The puzzle server shares a cryptographic key with the destination end host. The puzzle server can
use the end host's cryptographic key to derive a set of the end host's active IP addresses at
any instance of time (as discussed in \S~\ref{sec:destination}).

\paragraphb{Computational puzzles:}
When a source end host first contacts the puzzle server, it responds with a computational puzzle.
Upon receiving a solution to the computational puzzle from the requesting node, the server
sends back \prateek{an ephemeral active} %
IP address of the destination end host. We use Portcullis~\cite{parno:sigcomm07}
to implement the computational puzzles aspect.

\subsection{Source end hosts}
\label{sec:source}

The source end hosts first contact the DNS server to resolve the hostname for the
destination end host. If there is no DDoS attack on the destination, they receive
the conventional static non-hopping IP address for the destination. On the other
hand, the source end hosts are redirected to a puzzle server during a DDoS attack.
The source end hosts then contact the puzzle server to obtain \prateek{an ephemeral active} IP address
for the destination. The puzzle server first asks it to solve a computational puzzle.
The source end host solves the computational puzzle and sends it to the puzzle
server. The puzzle server returns an ephemeral \prateek{active} IP address of the destination,
which can be used to receive service.
\prateeknew{Our JavaScript mechanism in \S~\ref{sec:endhosts} enables source end hosts to access any 
Mirage protected web server without installing any software.}

\section{Deployment Details}
\label{sec:deployment}

Next, we describe %
extensions to the basic Mirage design that
simplify deployment and enhance \prateeknew{security/performance}.%
%

\subsection{Puzzle Server}
\label{sec:puzzleserver}

\begin{figure}
\centering
\includegraphics[width=2.4in]{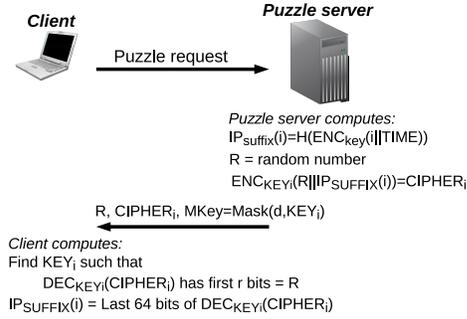}
\begin{security}
\vspace{-0.1in}
\end{security}
\caption{\prateeknew{Puzzle construction: active IP address of the server is the solution.}
}
\begin{security}
\vspace{-0.2in}
\end{security}
\label{fig:puzzle}
\end{figure}

\paragraphb{Eliminating puzzle verification:}
In the Mirage design described so far, the puzzle server issues the computational
puzzle seed to the clients, verifies that the clients have solved the
puzzle correctly, and sends \prateek{an active} IP address to the clients.
We can simplify the complexity
at the puzzle servers by having the \prateek{active} IP addresses be the solution to
the computational puzzles. This way, the puzzle servers act purely as computational
puzzle seed generation and distribution servers. Since IP addresses are puzzle solutions,
notice that the destination end-host must be able to efficiently compute the puzzle solutions, since
it would need to put the results of the computational puzzle in the network ACLs.
We term this feature to be a {\em trapdoor computational puzzle}. We now provide
constructions for such a trapdoor puzzle.
As before, let us assume that the destination end host and the puzzle server share a key. The destination end
host and the puzzle server can compute the set of active IP addresses
as before, i.e.,
\[ IP(i) = PREFIX || H( ENC_{KEY}(i || TIME)) \]

\prateeknew{Let us denote the first $64$ bits of the destination end host's IP address as the $PREFIX$.
 We denote the last $64$ bits of $IP(i)$ as $IP_{SUFFIX}(i)$ (equal to $H(ENC_{KEY}(i||TIME))$).
Let $KEY_i$ be a symmetric key. Let $R$ be a random number. Now, the puzzle server
first computes the following:}

\[ ENC_{KEY_i} (R || IP_{SUFFIX}(i)) = CIPHER_i \]

\prateeknew{Next, the puzzle server issues the following to the client a) the $PREFIX$, 
the $r$ bit random number $R$, the cipher-text of the encryption $CIPHER_i$, 
all but the last $d$ bits of $KEY_i$, where $d$ is the difficulty level of the puzzle.
The job of the client is to iterate through each of the $2^d$ possible values of $KEY_i$,
decrypt the cipher-text $CIPHER_i$, and check if the first $r$ bits
of the decryption match the random number $R$. Upon finding a match, the client can uncover a single
IP address by appending the last $64$ bits of the decryption to the $PREFIX$. Figure~\ref{fig:puzzle} illustrates
this computational puzzle design.}
\paragraphb{\prateeknew{Honest but curious puzzle servers:}}
We now extend Mirage to use the puzzle servers as untrusted 
\prateeknew{(honest but curious)} puzzle distribution servers.~\footnote{\prateeknew{Active attacks by puzzle servers, such 
as returning incorrect JavaScript to source end hosts, can be detected 
by the victim server (by anonymous querying), and the malicious puzzle servers 
can be blacklisted.}}
 The main idea is that the destination end host can communicate the puzzles
directly to the puzzle server, so that puzzle server does not have to generate
the puzzle. Moreover, by using the previous extension (\S~\ref{sec:puzzleserver}) where
the \prateek{active} IP address is the solution to the puzzle, the puzzle server does
not need to do any verification, eliminating the need for cryptography at the
puzzle server. The destination end host can simply chose a random set of \prateek{active} IP
addresses, generate the puzzles as in the previous extension, and store them at
the puzzle server. In this fashion, the puzzle server does not need to be
trusted with the destination end host's \prateek{active} IP addresses.  Note that a malicious
puzzle server could share the puzzles with the attackers ahead of time,
\prateekccs{enabling %
precomputation attacks.}
To mitigate this
attack, we propose that the destination end host upload the puzzles at the puzzle server every
hopping time interval.
\paragraphb{Eliminating puzzle server as a bottleneck:}
So far, we had assumed that the adversary could not launch a DDoS attack on the puzzle server.
Since the puzzle servers simply need to serve static puzzles and can be untrusted
\prateekccs{(due to above mechanisms)}, the destination end-host can ensure
availability of puzzle seeds by simple replication. For instance, the puzzles could be hosted
on Akamai %
\prateek{or can be widely replicated across cloud service providers such as Amazon S3.}
The adversary would need to DDoS all of the replicas in order
to DDoS the destination. Note that different replicas could either store different puzzles, or use
standard CDN approaches to return different puzzle objects to different source end hosts.

\subsection{End-hosts}
\label{sec:endhosts}

\paragraphb{Legacy source end hosts:}
We now discuss how source end hosts can take advantage of Mirage protected sites
without installing additional software.
\prateekccs{
When the source end host performs a DNS lookup for the destination end host, 
it is redirected to a DoS-resistant puzzle server (can be hosted by Akamai), 
which returns JavaScript code.} \footnote{\prateekccs{A small minority of users 
who visit the website in a short time window both before and after attack 
may be affected by browser DNS caching~\cite{jackson:ccs07}. We tested browser 
behavior for this scenario using Chrome/Firefox; we found that the first browser 
request (after attack) is timed out (cached IP address is unresponsive), 
but the subsequent reloads are successfully redirected to the puzzle server.
}}
\prateekccs{
The JavaScript solves the computational puzzle to obtain an active IP
address, and issues a cross origin request to that active IP address,
either using an \emph{iframe} or a \emph{XMLHttpRequest}.
The victim server can
set \emph{Access-Control-Allow-Origin} to its own domain, to allow 
such a cross origin request from the trusted JavaScript, and 
maintain functionality under the constraints of the browsers 
same-origin policy (SOP).
Sites that disallow 
framing for defending against clickjacking can set \emph{Frame-Options: Allow-From:} 
to their own domain. 
To enable subsequent requests from the client, the JavaScript can 
set the \emph{base} tag of relative reference links to the active 
IP address.} In this fashion, any standards compliant browser can 
view a Mirage protected site. Clients that don't use JavaScript 
can be directed to a best effort service.
\prateek{The ability to access Mirage protected websites without installing
any software makes our design most suitable for web applications.
}

\paragraphb{Achieving per-computation fairness:}
We note that in the protocol described in \S~\ref{sec:protocol},
honest source end hosts only solve a single computational puzzle per flow. This
may not result in per computation fairness, since the honest node's
computational resources may be under-utilized.  We mitigate this problem as
follows: when the honest source end hosts obtain \prateek{an active} IP address for the
destination end host (after solving a computational puzzle), they continue to
spend additional computational resources, which can be used to derive new
\prateek{active} destination IP addresses (by querying for and solving another puzzle).
This modification in our protocol results in per-computation fair allocation of
resources.  %
\prateek{We now discuss how source end hosts can load balance their traffic over all
available active IP addresses.}
To switch between available \prateek{active} IP addresses, %
the JavaScript mechanism discussed above could fetch each image or object in the
page via a different \prateek{active} IP address. 
\prateekccs{For larger objects, the JavaScript could make multiple requests 
for particular \emph{byte ranges} of the content (using HTTP range request header),
and assemble the returned contents to form the requested page.}

\paragraphb{Handling hopping interval transitions:}
\prateek{
In Mirage, we set the hopping interval to 5 minutes, as discussed in \S~\ref{sec:security}.
A majority of web traffic flows in the Internet are short lived~\cite{short-lived-flows}, and will finish
within a single hopping interval. For scenarios where a short-lived flow starts close to the hopping
point, and for long lived flows, observe that the JavaScript is already solving computational puzzles
continuously, to achieve per computation fairness. Based on our assumption of loose time synchronization
amongst end hosts, the JavaScript can start to solve the puzzles for the next time period and thus
receive uninterrupted service.
To prevent unnecessary loss in this scenario, the destination continues to receive traffic on the old
set of IP addresses (from the previous hopping interval) for a time threshold. %
\prateekccs{Finally, we note that all cross origin requests are made on IP addresses, 
and thus DNS caching by browsers does not impede JavaScript's ability to hop IP addresses.} 
}

\subsection{Network}

\paragraphb{Filtering traffic:}
The destination (victim) end host's upstream routers need to filter traffic to
all non-active IP addresses for the victim. We propose two strategies for
filtering traffic, %
\prateek{both of which use existing router interfaces.} The first strategy is
to use an IRSCP~\cite{irscp}-like management server within the destination's
upstream AS. The destination end host could communicate its list of
\prateek{active} IP addresses to the management server, and the management
server could instruct routers to drop traffic to the remaining non-active IP
addresses via iBGP routing updates\footnote{Note that iBGP routing updates are
local to the AS and do not cause routing instability.}.
Alternatively, the management server could push
configuration files containing ACLs for filtered IP addresses to routers.
\prateek{We now discuss how big the ACL \matt{should be} to support our protocol.} Suppose that the
local network is under a DDoS attack, and that network supports hundreds of
thousands of users within the time duration of the hopping interval
(set to 5 minutes), typical of \matt{large} data center environments~\cite{devoflow}.
Also suppose that we are interested in defending against a botnet with 100\,000 bots.
In this scenario, it suffices to install an ACL with a few hundred thousand entries (estimated using prior usage history)
at the victim's upstream ISP (assuming that it is the bottleneck). Current routers
can already support millions of ACL entries~\cite{nanog39}, and would not need to be
upgraded in this scenario. On the other hand, to defend against larger botnets,
or when the deployment happens further upstream, the ACL size could reach tens of
millions of entries, in which case existing filtering mechanisms may need to be
scaled \matt{(which the operator can do by installing more memory at the router, rather than requiring cooperation from the router vendor)}.
\begin{security}
\prateek{We propose an adaptive version of Mirage which can further reduce filter table size in Appendix~\ref{sec:discussion}.}
\end{security}
\begin{techreport}
\prateek{We propose an adaptive version of Mirage which can further reduce filter table size in \S~\ref{sec:discussion}.}
\end{techreport}
\paragraphb{Fair sharing:}
Routers already support per-destination fair queueing, and this option can also be manipulated by
a configuration files pushed by a management device.
For example, the Cisco secure policy manager~\cite{ciscosecurepolicyman}
can read in high level description of policies for a network, and
translate them into low level specifications. When a destination
end host is under a DDoS attack, it can request its upstream AS to enable per-destination
fair queueing, who can then automatically publish the corresponding configurations to
routers. Alternatively, if desired, the upstream AS could
eliminate the need for a policy manager device by leaving fair queuing always enabled for the customer (some ISPs already run similar QoS mechanisms
to improve service for their customers and their own traffic).
\prateekccs{Observe that the state required to support per destination fair queueing 
at an upstream router only depends on the number of flows towards downstream 
destination IP addresses in a short time interval (and not the total downstream IP address space).} 
Finally, we note that our design does not require perfect fair queueing. Thus various
traffic monitors located in the upstream AS could check if some \prateek{active} IP addresses are receiving
more traffic than others, and if so, then push either a iBGP routing update or an ACL update to
block those \prateek{active} IP addresses. 
\paragraphb{Supporting IPv4:}
Mirage is ideally suited for IPv6, given IPv6's large address space. While
implementing our architecture in IPv4 has some challenges, it is feasible.  We
consider multiple scenarios for IPv4. In the first scenario, the source end
host supports only IPv4, but the destination server and its upstream provider
(which is most likely to do filtering) support IPv6.  In this case, the source
end hosts need to run an IPv4-IPv6 translator, but the network continues to
function as before.%
We note
that Teredo IPv6 tunneling~\cite{teredo} is built into Windows and is available
on Linux and Mac OS X.
In the second scenario, the source end host supports only IPv4 and the
destination end host's network provider also supports only IPv4. %
In this scenario, the destination end host can set up a tunnel to another
remote IPv6 provider. To prevent attackers from targeting its IPv4 address, the
destination end host can request its local provider to filter all access to its
IPv4 address (except from its IPv6 tunnel). Now, the victim can advertise its
IPv6 address and perform IP address hopping as before. As in the above
scenario, source end hosts use an IPv4-IPv6 translator, and the network does
not need any software or hardware updates.  Alternatively, the victim server
could purchase service directly from an IPv6-enabled provider.

\section{Security Analysis}
\label{sec:security}

Next, we analytically evaluate the security of our design. %

\paragraphb{Network scanning attacks:} A possible threat to Mirage is the use
of network scanning attacks to uncover the set of active IP addresses in use by
an end host. Due to the large range of possible IP addresses for
an end host ($64$ bits), such attacks are  already quite costly. The set of
active IP addresses changes after every hopping time interval (which we set to
$5$ minutes); thus the attacker has to scan the entire range within the
hopping time interval.
Consider two attack scenarios: (a) the size of botnet owned by the attacker is
$20,000$ (the average size of modern botnets~\cite{paulson:comp06}) and
(b) the size
of the botnet is $1$ million nodes (an extreme scenario).  Now, if the bots
have $1$Mbps links to their provider, each bot would be able to send only about
$2^{25}$ ping packets over the 5 minute interval. Thus, in the two attack
scenarios, the attacker would only be able to scan an insignificant fraction
of the range of IP addresses for an end host - $2^{-28}$ and $2^{-23}$
respectively.

\paragraphb{Brute-force DoS attacks:} The attacker could simply send attack
traffic to the victim using a random interface (subnet) address for each packet, with
the idea that a fraction of its traffic would be able to bypass router
filters and congest the victim link. With perfect filtering of non-active IP
addresses at the routers, such brute-force DoS attacks would be completely
ineffective as the probability of a random IP address being an active IP
address is very small. E.g., suppose there are about $5000$
client nodes per $5$ minute interval per destination end host. Then, the set of
active IP addresses needed to contain about $25000$ and $1,005,000$ entries
respectively (for the two attack scenarios discussed above), resulting in the
success rate of a random attack of only $1.3 \cdot 10^{-15}$ and $5.4 \cdot
10^{-14}$ respectively.

\paragraphb{Theoretical Results:} Next, we present key analytical results which illustrate 
that (a) Mirage's computational puzzle mechanism is secure, (b) Mirage provides per computation
fairness, and (c) Mirage is able to gracefully deal with compromised routers. 
{\theorem \label{thm:puzzle-sec} Puzzle scheme security: 
For a computational puzzle with difficulty level $d$, 
the attacker has negligible advantage in
obtaining \prateek{active} IP addresses compared to honest sources.
}
\proof{
\prateeknew{We can model the AES block cipher as a pseudorandom permutation. If the attacker has
any advantage over an honest source end host's brute force strategy for solving the computational puzzle,
then the attacker would be able to break the pseudorandom permutation as well. Thus, the best strategy
for the attackers is a brute force through the $2^d$ possible keys, which is identical to the honest client
strategy.}}
{\theorem \label{thm:fairness} Per computation fairness: 
Let $C_A$ and $C_H$ denote the computational resources for the attacker and honest sources
respectively. The honest source end host's share of the victim bandwidth is $\frac{C_H}{C_H+C_A}$ }
\proof{
From Theorem~\ref{thm:puzzle-sec}, we know that the attacker's advantage
over honest sources in solving the computational puzzles is negligible.
We have already shown that network scanning attacks to uncover \prateek{active} IP addresses are ineffective.
Also, from Section~\ref{sec:endhosts}, source end hosts fully utilize
their computational resources by continuing to solve computational
puzzles even after obtaining a single \prateek{active} IP address for
the destination. Thus after any amount of time $t$, the number of \prateek{active} IP
addresses known by the attackers and the honest clients are proportional to
$C_A$ and $C_H$ respectively. Finally, in our protocol, the network performs
per destination fair queuing, resulting in fair sharing of resources amongst \prateek{active} IP addresses.
Thus the honest source end hosts get $\frac{C_H}{C_H+C_A}$ of the total bandwidth.
}
{\theorem \label{thm:routers} Impact of compromised routers: 
Let a set of compromised routers carry a fraction $f$ of the legitimate traffic towards a
destination end host under attack. If the set of compromised routers collude with the adversary,
then the honest sources' %
share of the victim bandwidth is $\frac{C_H \cdot (1-f)}{C_H + C_A}$
of the total bandwidth.}
\proof{In Mirage, malicious routers that handle a fraction $f$ of the legitimate traffic
towards the destination end host can snoop on the legitimate traffic towards the
destination end host and uncover its \prateek{active} IP addresses. Moreover, in addition
to snooping on the legitimate traffic, the malicious routers can drop the legitimate
client traffic towards the destination to increase the attacker's bandwidth share.
Thus the attackers will be able to obtain a fraction $\frac{C_A+f\cdot C_H}{C_A+C_H}$ of the
total bandwidth, while the honest source end hosts obtain a fraction
$\frac{C_H \cdot (1-f)}{C_H + C_A}$ of the total bandwidth.
\footnote{We note that compromised routers in the destination's upstream AS also
have the potential to reveal the access control list to the adversary, but
the upstream AS can easily detect and isolate such compromised routers by inserting
a few spurious entries in the ACL and checking for traffic towards those destination
IP addresses.}}
\begin{techreport}
We discuss the issue of insider collusion attacks in Section~\ref{sec:discussion}.
\end{techreport}

\section{Evaluation}
\label{sec:evaluation}

In this section, we quantify the attack resilience of Mirage using a
prototype implementation and an ns-2 simulation, compare
Mirage with other DoS defenses, and perform Internet
bandwidth measurements to determine the extent to which Mirage must
be deployed in order to provide protection against various levels of
DoS.

\subsection{Prototype Implementation} \label{sec:prototype_impl}

To evaluate Mirage's performance and deployability, %
we built a prototype implementation of Mirage.
We did not attempt to optimize
the prototype system; rather, our goal is to verify the design and evaluate its
deployability. To evaluate our prototype, we constructed a {\em Mirage-enabled
web service}, which we call the \emph{victim server}, and configured a set of
clients to access that web service.

\begin{figure*}[!t]
\centerline{ \subfigure[Prototype implementation architecture.]
{\includegraphics[width=0.30\textwidth]{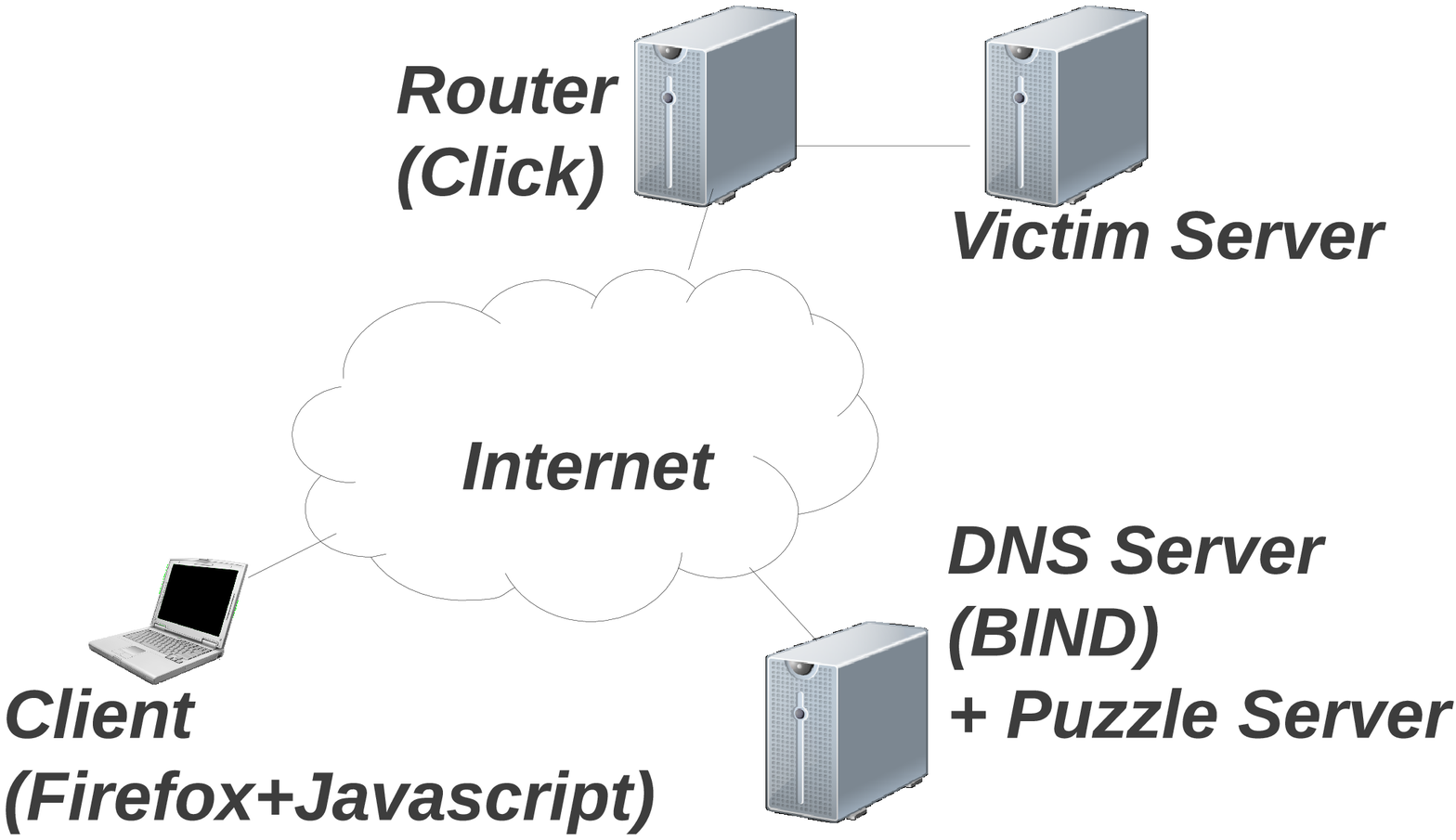}\label{fig:prototype_impl}}
\hspace{4pt} \subfigure[Experimental setup.
]{\includegraphics[width=0.30\textwidth]{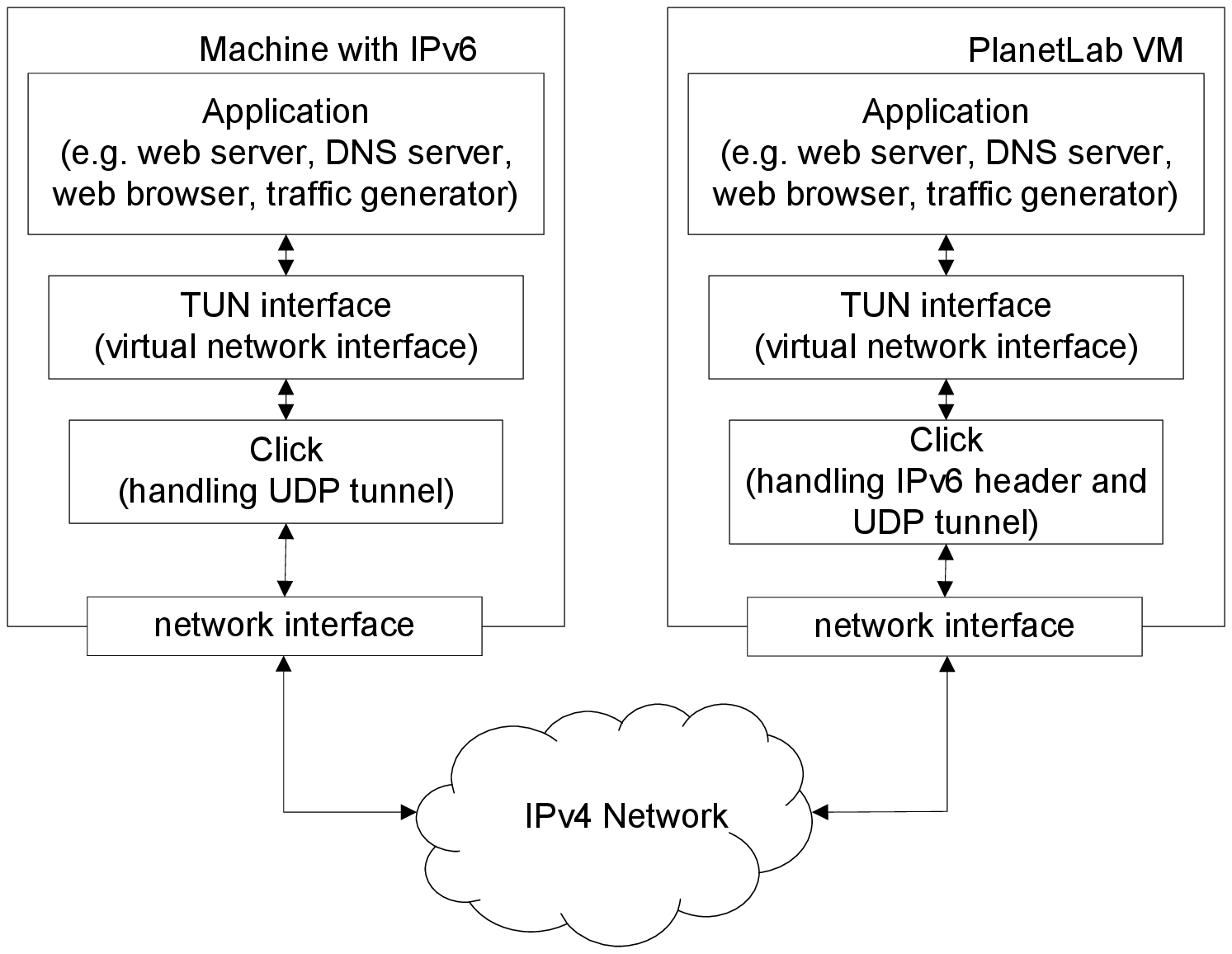}\label{fig:planetlab_exp}}
\hspace{4pt} \subfigure[Mirage guarantees normal users' throughput even with high-rate attackers.
]{\includegraphics[width=0.30\textwidth]{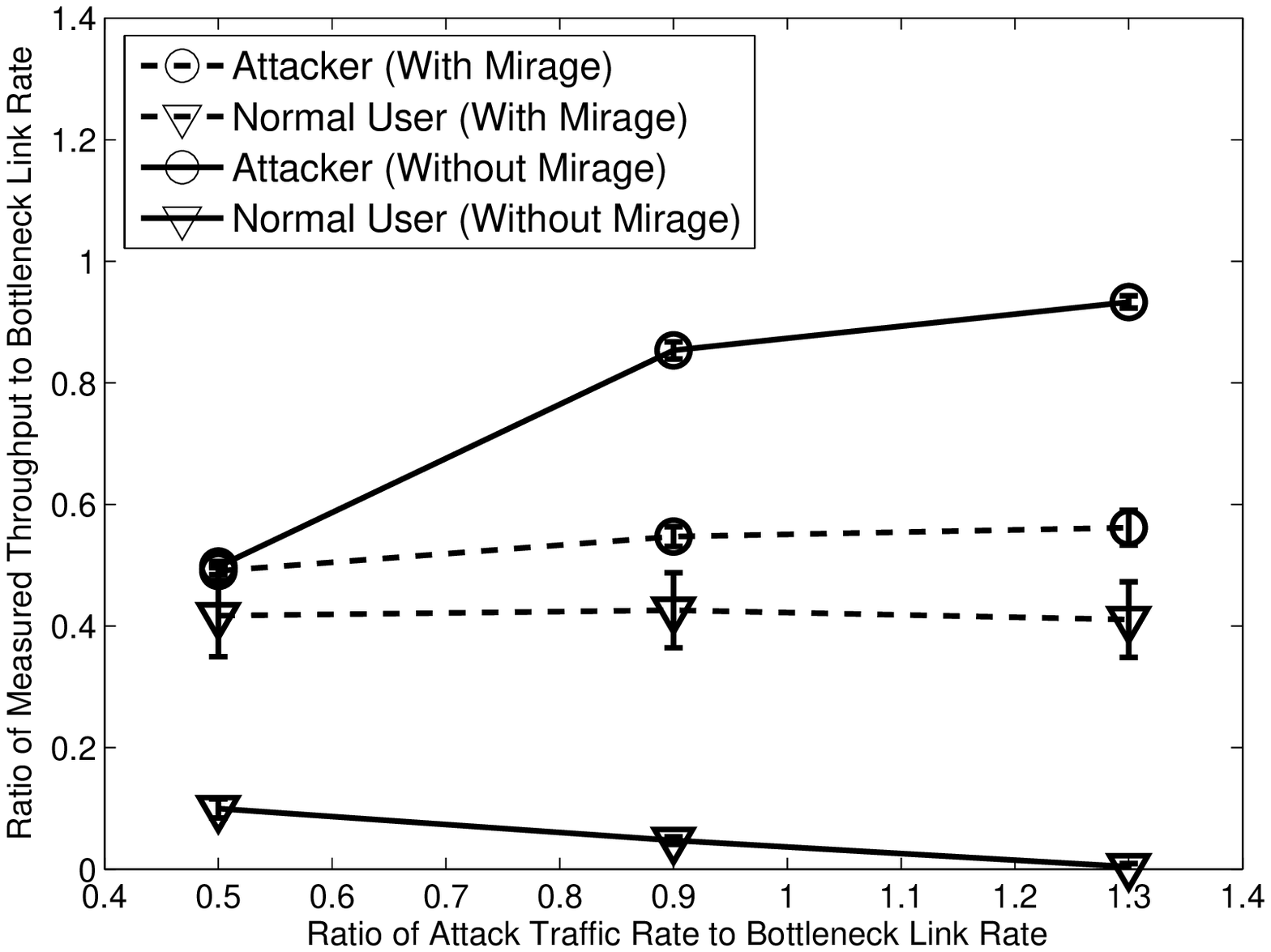}\label{fig:fair_queueing_evaluation}}
}\vspace{-9pt}
\begin{security}
\vspace{-0.1in}
\end{security}
\caption{Prototype implementation.} \vspace{-10pt}
\label{fig:implfigs}
\vspace{-0.1in}
\end{figure*}

\paragraphb{Client side:} We implemented the JavaScript program described in
\S~\ref{sec:endhosts} to enable the client to ``hop''
addresses without requiring changes to the client's browser or operating system. %
This JavaScript %
is located at the puzzle server
as shown in Figure~\ref{fig:prototype_impl}.  When the DNS directs a client to
the puzzle server, the client fetches and solves the puzzle by running the
JavaScript %
provided by the puzzle server.  These operations are
transparent to the user; the user simply directs its %
browser to %
the victim's domain name.

\paragraphb{Server side:} The server-side components, comprised of the victim
server, router, DNS server, and puzzle server, provide \emph{transparent}
protection for a victim server.  By transparent, we mean that the user of a
client machine need not install any additional software.
We used BIND~\cite{tool:bind} to implement the DNS server. We
defined our own zone (miragev6.org) and created a probing tool that
regularly sends pings to the victim server to detect when the victim
server 
is under attack. Initially, www.mirage v6.org resolves
directly to the victim's IP address. Once the probing tool fails to
receive a certain number of echo responses, it changes the DNS
record so that www.miragev6.org resolves to the puzzle server's IP
address (which, in our prototype, is the DNS server). We set the TTL
for the DNS record to a small value so that an entry cached before
the victim was under attack will be quickly corrected %
after the attack.
When the victim is under DoS attack and
the client tries to access www.miragev6.org,
the client receives the IP address of the puzzle server.
The client connects to the puzzle server and fetches %
JavaScript code which includes a puzzle.
Our implementation uses the scheme %
in \S~\ref{sec:puzzleserver},
using AES as the encryption algorithm.

\paragraphb{Experimental setup:}
To investigate Mirage's operation in the wide-area, we conducted
experiments in the Internet using the PlanetLab testbed.
Figure~\ref{fig:planetlab_exp} shows how we set up the network. We
selected random (lightly CPU-loaded) nodes to run our experiments.
We selected one PlanetLab node to act as the client, and one
PlanetLab node to run the server, router, puzzler server, and DNS
software. Since PlanetLab hosts do not directly provide IPv6
compatibility, for our IPv6 experiments we used an IPv4 encapsulated
tunnel. We found that standard tunnel (e.g. GRE) packets are blocked
by firewalls of some PlanetLab networks, so we used UDP tunnels for
our experiments. We implemented this tunnel with the
TUN~\cite{tool:tuntap} virtual interface and
Click~\cite{tool:click}. In an IPv6-enabled machine, outgoing IPv6
packets are delivered to TUN device which lets Click capture the
packets. Click then encapsulates the packets with UDP tunnel.
Because PlanetLab nodes lack kernel support for IPv6, we assigned
IPv4 addresses to the TUN device and used Click to translate between
IPv4 and IPv6 addresses.

\subsection{Performance Evaluation}

In this section, we evaluate how well Mirage can defend against
possible attacks. We do not argue that we consider all possible
attacks; rather, our intention is to quantify Mirage's attack
resilience.
We consider two attacks: one that exhausts network
bandwidth and one that exhausts hopping IP addresses.

\paragraphb{Bandwidth exhaustion attack:}
By increasing its sending rate,
or %
the number of flows it sends,
an attacker can increase its traffic rate
to overwhelm normal users' traffic.
We examined the effect of the attack traffic rate on benign traffic
by setting up
ten benign TCP traffic flows and one UDP attack traffic flow
that share a single bottleneck link to the victim.
We measured the throughput of the benign and the attack traffic,
and compared them %
with and without Mirage.
Figure~\ref{fig:fair_queueing_evaluation} shows the results
when we vary the attack traffic rate between 0.5 to 1.3 times
the bandwidth of the bottleneck link.
Without Mirage, as the attack traffic rate increases,
the normal user's traffic rate decreases;
with Mirage,
fair queueing drops the attackers' excess packets
rather than the normal user's traffic.

\begin{figure*}[!t]
\centerline{ \subfigure[Implementation study with a small number of
attacker processes (3 attackers)
]{\includegraphics[width=0.30\textwidth]{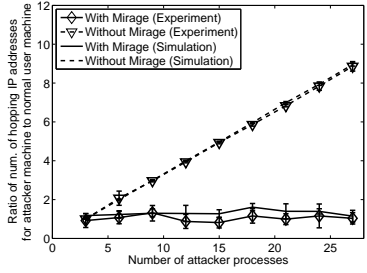}\label{fig:puzzle_evaluation_small_scale_2}}
\hspace{4pt} \subfigure[Simulation study with a large number of
attacker processes (3 attackers)
]{\includegraphics[width=0.30\textwidth]{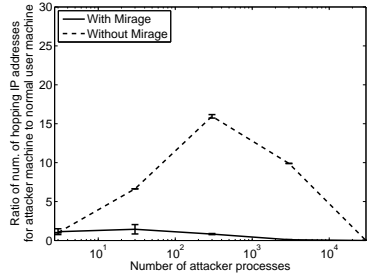}\label{fig:puzzle_evaluation_machine_3_2}}
\hspace{4pt} \subfigure[Simulation study with a large number of
attacker processes (30 attackers)
]{\includegraphics[width=0.30\textwidth]{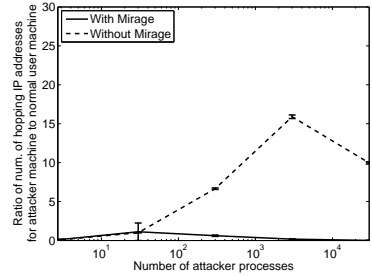}\label{fig:puzzle_evaluation_machine_30_2}}
}\vspace{-9pt} \caption{Defense against hopping IP address
exhaustion attack.}\vspace{-10pt}
\label{fig:puzzle_evaluation_2}
\vspace{-0.1in}
\end{figure*}

\paragraphb{Hopping IP address exhaustion attack:}
Since Mirage uses fair queueing per hopping IP address, an attacker
can request a large number of hopping IP addresses to prevent normal
users from getting hopping IP addresses, thus preventing normal
users from transmitting to the victim. In order to evaluate Mirage's
performance under a large number of attackers, we use the ns-2
simulator~\cite{simulator:ns}. To parametrize the
simulation, we developed a computational model and used Internet
experiments using a small number of attacker processes to estimate
the parameters for our model.

Our Internet experiments used three machines to behave as an
attacker and one machine to behave as the normal user. We
implemented a client using C code to access the
Mirage-enabled web service. To demonstrate the performance
of our JavaScript extension, we also considered Firefox
to be the client. Both normal users and attackers solve
puzzles as quickly as possible in order to obtain hopping addresses;
however, in normal user machines, a client process generates IP
addresses, whereas on attacker machines, we vary the number of
client processes that are generating IP addresses.
In our prototype implementation, the puzzle
server uses hopping IP addresses released from a leaky bucket and
assigns it to the puzzle solution with the highest difficulty.
A process that is not chosen will increase its puzzle difficulty and
try again. Without Mirage, the puzzle server grants all requests for
IP addresses. We measured the number of hopping IP addresses
obtained by the attacker machines and the normal user machine.
Figure~\ref{fig:puzzle_evaluation_2}a
shows the ratio of
the number of hopping IP addresses per attacker machine to the
number of hopping IP addresses of normal user machine when we use
Firefox as a client. Without Mirage, as the
number of attacker processes increases, the attacker machines can
get a larger number of hopping IP addresses, which corresponds to a
large number of victim machine accesses. Note that for Firefox
processes, the growth rate becomes smaller as the number of
processes increases because the additional processes will saturate
the machine's computational power. We can see that with C code, the
saturation of computation power is reduced. With Mirage, this
computational limit is reached much more quickly, because a node's
ability to get IP addresses is computationally limited.

We used this experimental data to establish a model in ns-2. We used
a dumbbell topology with RTT equal to that from our experimental
study. When a node solves a puzzle solution, it takes an amount of
time to generate a solution given by
$\frac{B+N \times E }{C}$
where $B$ is the cost (in cycles) of running a client process
without puzzle calculations, $N$ represents the number of encryption
attempts needed to solve a puzzle, which we model as a normal
distribution, $E$ is the number of cycles taken by each encryption
attempt, and $C$ is the speed of the CPU.
Without Mirage, a node takes time $B/C$ to
resolve the victim's IP address.
We varied the values of $B$, $N$, and $E$ to find parameters
that best matched our experimental study;
that is, ones that minimized the variance
between experimental and simulation results.
Figure~\ref{fig:puzzle_evaluation_2}a shows how well the
two sets of data match.
With these simulation parameters, we performed simulation studies to
show how the number of attacker processes affects the number of
hopping IP addresses that the attacker machine obtained.
Figures~\ref{fig:puzzle_evaluation_2}b and~\ref{fig:puzzle_evaluation_2}c
show the results when
the number of attacking machines is 3 and 30, respectively. We vary
the number of attacker processes from 3 to 30000. In all cases, we
can see that with Mirage, an attacker cannot obtain as many hopping
IP addresses as in the case without Mirage. Note that the attacker
performance dips beyond a threshold number of attacker processes due
to resource constraints at the attacker machines.

\subsection{Comparative Study}

\begin{figure}[!t]
\centerline{ \subfigure[File transfer completion]
{\includegraphics[width=0.24\textwidth]{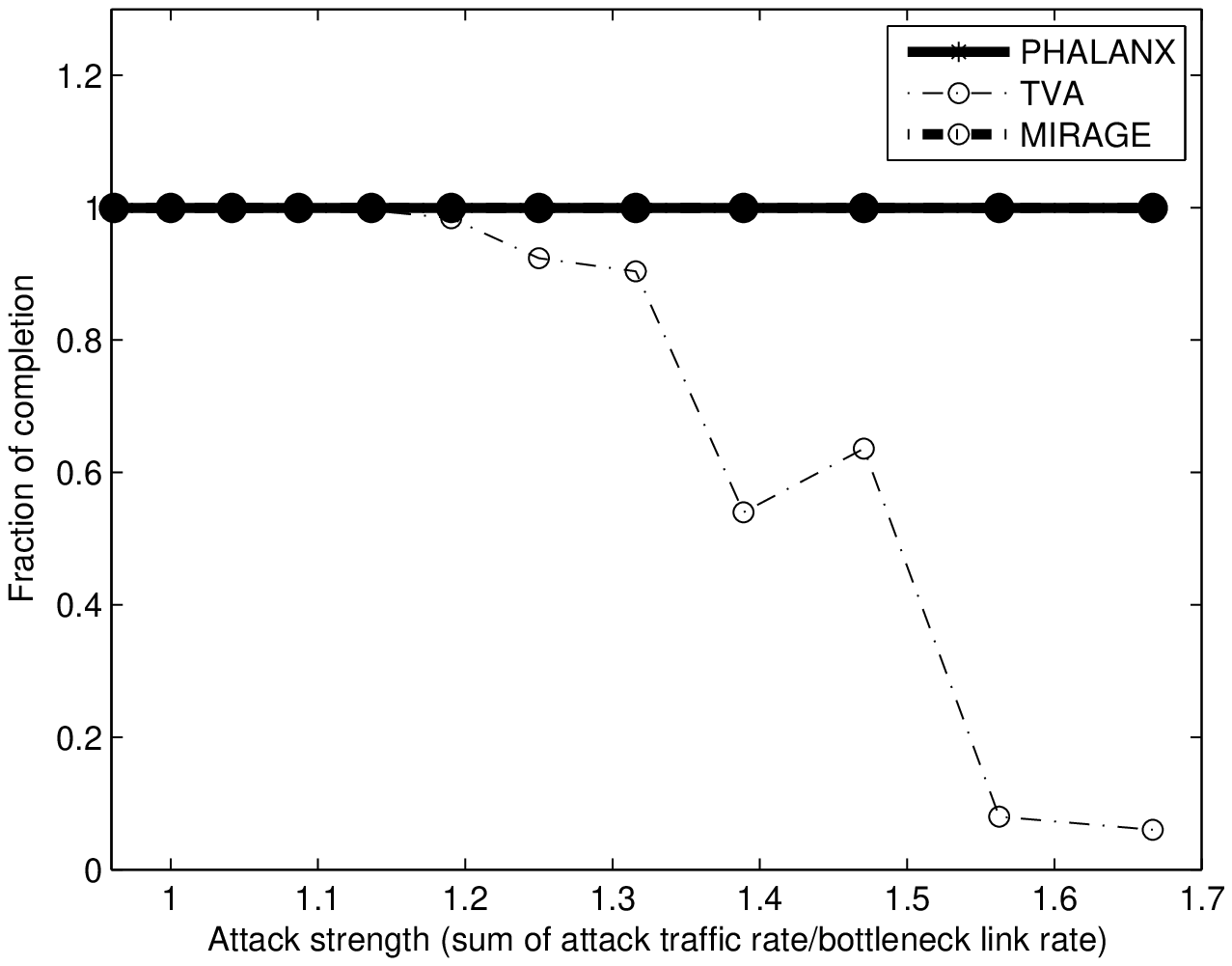}\label{fig:comparison_fraction}}
\hspace{4pt} \subfigure[Average transfer
time]{\includegraphics[width=0.24\textwidth]{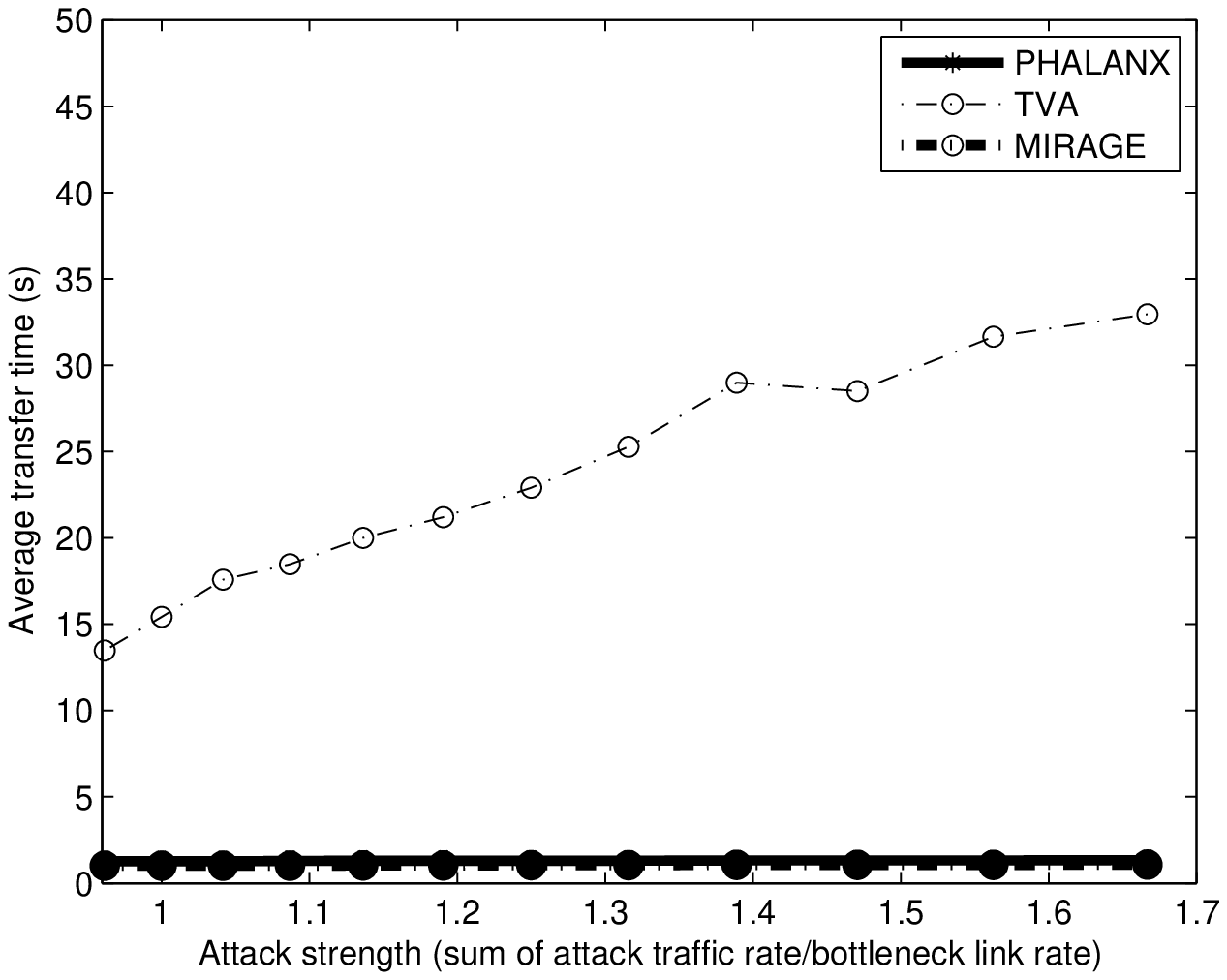}\label{fig:comparison_avg_transfer}}
}\vspace{-9pt} \caption{Comparative study varying %
link
bandwidth} \vspace{-10pt} \label{fig:comparison}
\vspace{-0.05in}
\end{figure}

\begin{figure}[!t]
\centerline{ \subfigure[File transfer completion]
{\includegraphics[width=0.24\textwidth]{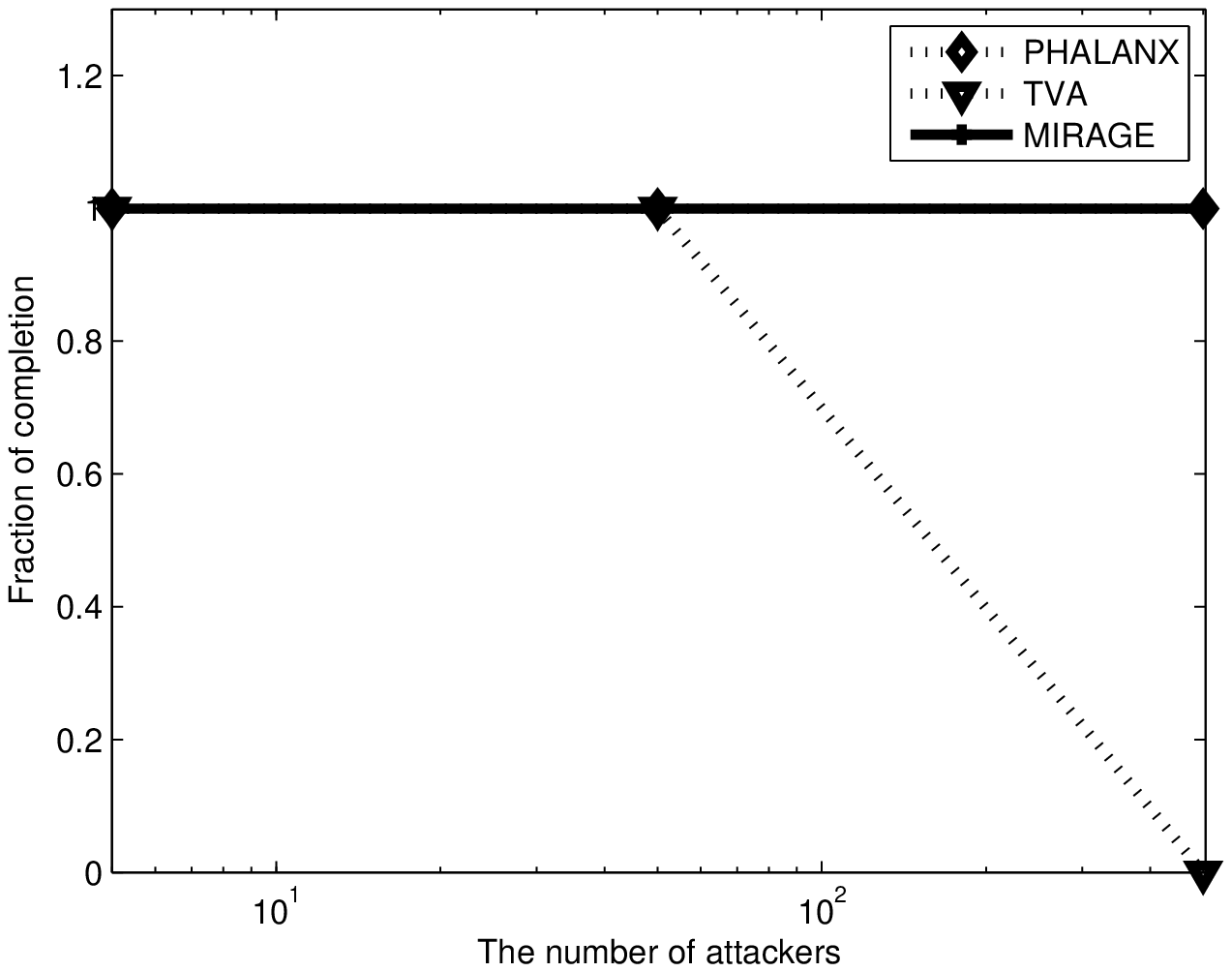}\label{fig:comparison_fraction_varying_num_attacker}}
\hspace{4pt} \subfigure[Average transfer
time]{\includegraphics[width=0.24\textwidth]{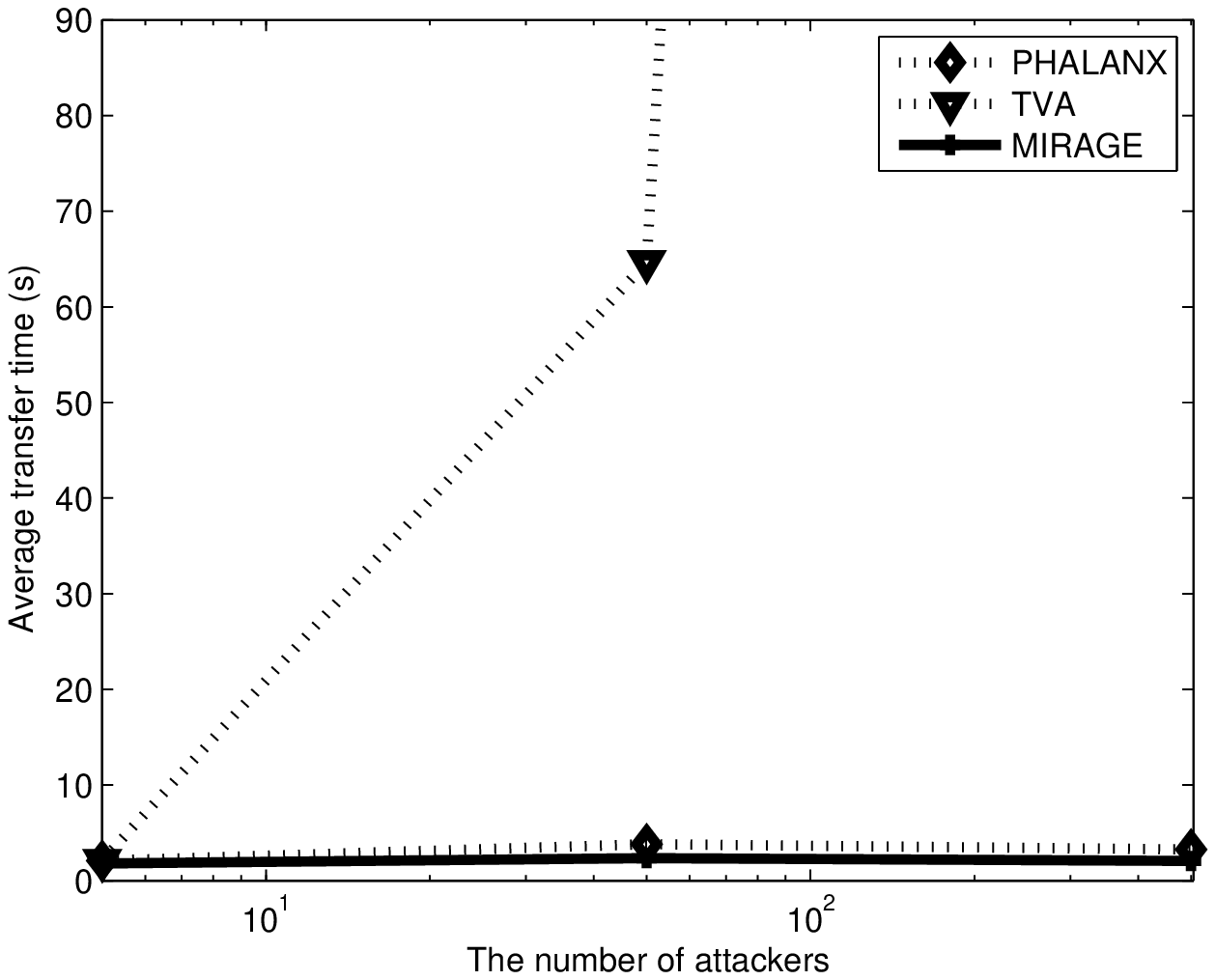}\label{fig:comparison_avg_transfer_varying_num_attacker}}
}\vspace{-9pt} \caption{Comparative study varying number of
attackers} \vspace{-10pt}
\label{fig:comparison_varying_num_attacker}
\vspace{-0.1in}
\end{figure}

\matt{ %
We compare the effectiveness of different DoS defense mechanisms:
Phalanx~\cite{dixon:nsdi08}, TVA~\cite{yang:sigcomm05} and Mirage,
using the ns-2 simulator. For Phalanx, we developed ns-2 code and
for TVA, we used the code developed by authors of
TVA~\cite{yang:sigcomm05}. For Mirage, we used DRR as our fair
queueing mechanism. We followed the methodology in the work of Liu
et al.~\cite{liu:sigcomm08}. We used the IPv4 Routed /24 Topology
Dataset~\cite{tool:ipv4_routed_24_topology} to derive the AS-level
topology, which has links annotated with round trip times at each
hop. We scaled down the AS-level topology by 1/200 since the memory
of our simulation machine does not support the full Internet-scale
AS-topology. In our simulated environment, ten legitimate TCP
senders try to send 50 files \prateek{each}, of size 2KB. Each
sender immediately starts sending a file after its previous file
transfer is done. \prateek{We} measured the fraction of successful
file transfers and the average time to deliver files. While
legitimate senders send TCP traffic, attackers send UDP traffic. All
legitimate and attacker traffic goes through the same bottleneck
link to a destination. For the first simulation, we intended 10K
attackers in the Internet-scale. Hence in our simulation scaled
down, the number of simulated attackers is 50, and each attacker
sends 10Kbps of UDP traffic.
Figure~\ref{fig:comparison} shows results when we vary the
bottleneck link bandwidth from 300Kbps to 520Kbps. For the second
simulation, we vary the number of attackers from 5 to 500 fixing the
bottleneck link bandwidth as 500Kbps.
Figure~\ref{fig:comparison_varying_num_attacker} shows the results
in this case. Overall, we find that Mirage achieves its deployment
benefits %
while providing comparable results to Phalanx and TVA across a variety of attack
strengths.}

\subsection{Bottleneck Measurements on PlanetLab}

\begin{figure*}[ht]
\centerline{ \subfigure[Hop count
]{\includegraphics[width=0.3\textwidth]{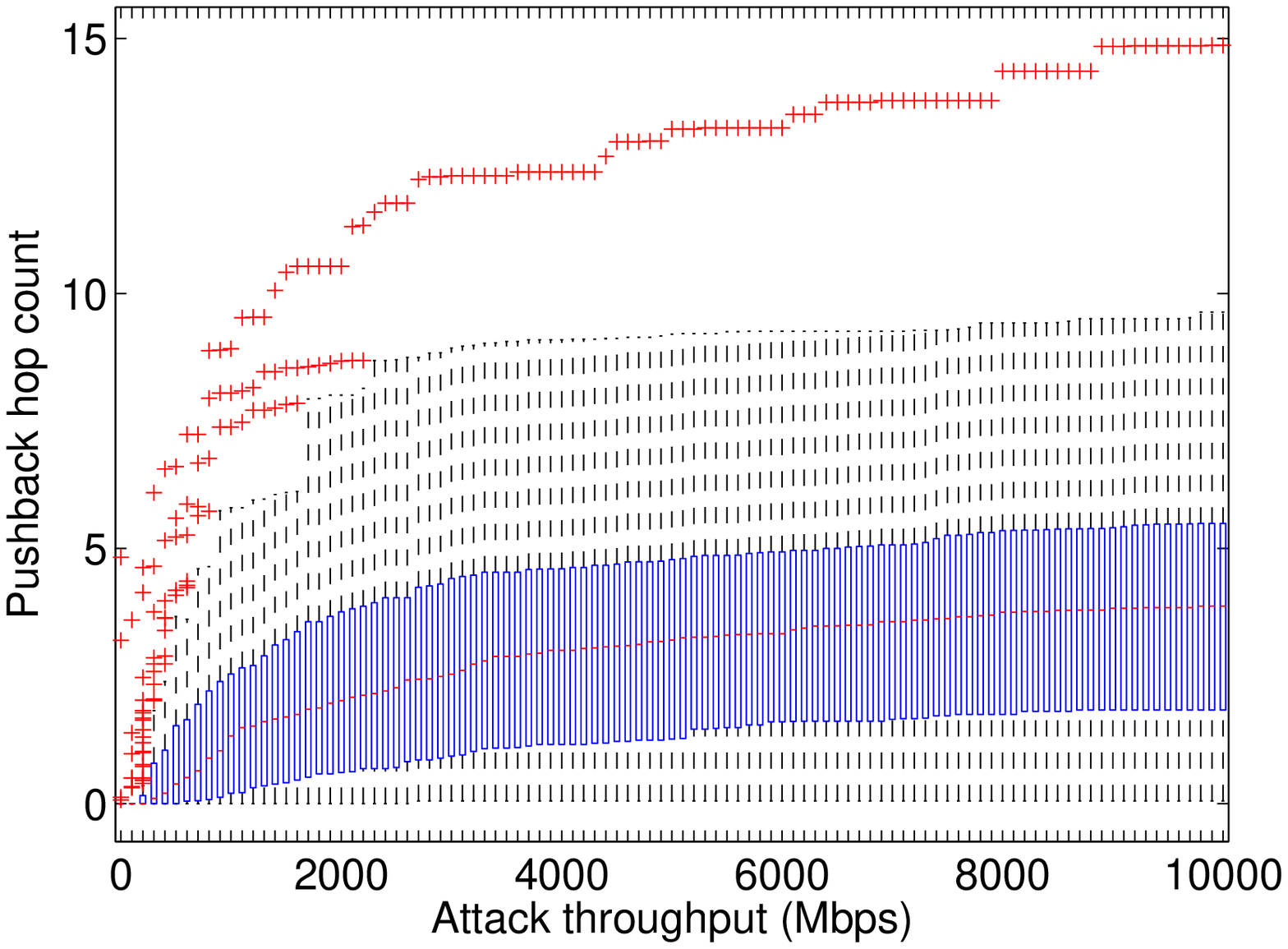}\label{fig:pushback_hop_count}}
\hspace{4pt} \subfigure[AS
]{\includegraphics[width=0.3\textwidth]{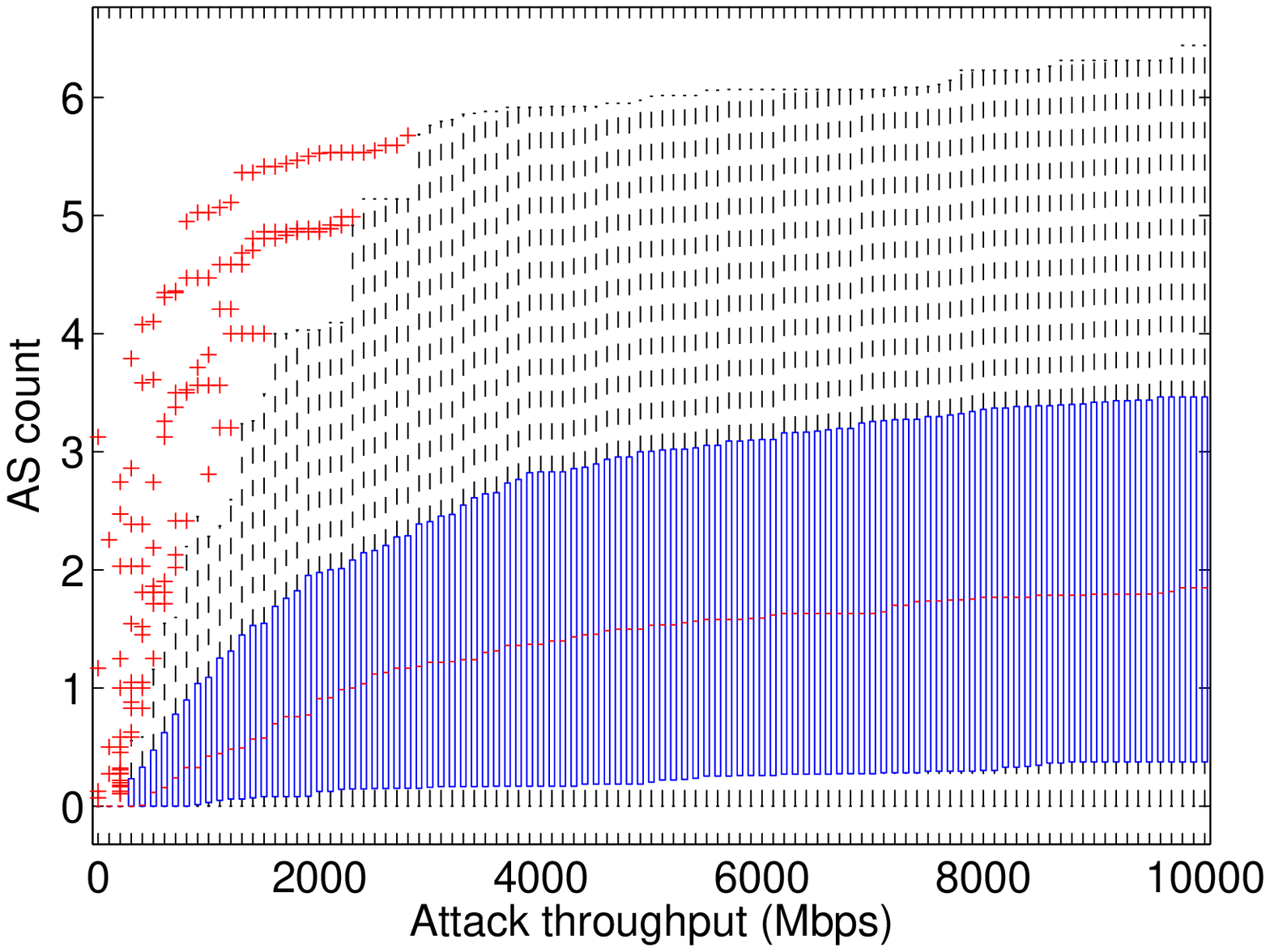}\label{fig:pushback_hop_count_as}}
}%
\begin{security}
\vspace{-0.1in}
\end{security}
\caption{Running Mirage at only the server's local network and ISP provider (first AS-level hop) is sufficient to achieve most of Mirage's benefits.}%
\label{fig:pushback_hop_count_results}
\begin{security}
\vspace{-0.2in}
\end{security}
\end{figure*}

The goal of this study is to compute the {\em pushback links}: the nearest
links to the victim server that are sufficient to block the attacker's traffic
while still letting non-attack traffic through. Computing these links sheds
light on how close to the network edge (router hops/AS hops) Mirage can be
deployed while still realizing its benefits.  
To gain information about link bandwidths in
today's Internet, we ran {\em pchar}, an open-source reimplementation of
pathchar, from $336$ randomly-selected PlanetLab nodes to the top 100 %
websites from alexa.com.
We constructed a target IP address set by selecting a single IP address
per nslookup query (the first IP address).  
We ran pchar using $32$ repetitions per hop, and testing each hop with
$300$ byte increments. We ignored route changes during the course of a
pchar experiment, by setting the -c flag.  
To improve accuracy of pchar data, we
also ran the {\em pathneck}~\cite{pathneck} bandwidth probing tool over the
same set of links.  Unlike pchar, pathneck measures the available bandwidth
rather than the capacity of links. Any capacity estimate lower than the
available bandwidth estimate was considered an error.  In addition, we combined
multiple experiments for improved accuracy.  For the links closer to the target
hop, we may have several pchar estimates from different PlanetLab nodes. In
this case, we can improve accuracy by considering the median value of the
estimates to be the final estimate of the link capacity (and thus mitigate the
effect of outliers).  If all pchar estimates for a link were %
erroneous, then for %
our analysis, we %
assume that such
hops are not the %
bottlenecks for the DDoS attack.

\paragraphb{Data analysis methodology:} 
For our analysis, we assume that the distribution of malicious
nodes in the network (bot nodes) is similar to the distribution of PlanetLab
nodes selected for our experiments.  While our assumption is only a heuristic,
our methodology is one way to shed light on the problem (without actually
carrying out a DDoS attack).
For each target IP, we first combine the results of all PlanetLab experiments,
and merge them into a single network map, rooted at the target IP. We annotate
links in the map with the fraction of paths traversing that link, and the
median of the capacity estimates.  Since we assume that the distribution of
PlanetLab nodes is similar to the distribution of malicious nodes, we first
approximate the amount of attack traffic traversing each link, for particular
attack throughput values. For example, if the total attack throughput is
$1$Gbps, and $10\%$ of the PlanetLab paths (probes from PlanetLab nodes to the
victim server) traverse that link, then the amount of attack traffic traversing
that link is $0.1*1$Gbps = $100$Mbps.  Using the bandwidth annotated network
map, it is now possible to compute the set of optimal pushback links. 
Finally we compute the weighted mean of the router level hop counts for the set of pushback links, where the weight 
indicates the fraction of attack traffic traversing that router. In addition to computing router-level hops for the 
bottleneck points, we also consider AS-level hops. For this purpose, we compute the IP to AS mapping using a BGP RIB snapshot collected from Routeviews~\cite{routeviews}.
Our study does not account for the scenarios where %
an AS has multiple 
AS numbers or when prefixes within a single AS number have multiple owners.

\paragraphb{Experimental Results:}
Figure~\ref{fig:pushback_hop_count_results} depicts the distance to pushback links as a function of the attack traffic using 
a boxplot%
, both in terms of the router and AS hop counts. When the attack traffic is less than $2$ Gbps \prateekccs{(vast 
majority of real world attacks in 2011~\cite{kaspersky})}, the median distance to the pushback links 
is less than 2 router hops and 1 AS hop. Thus running Mirage at only the victim's local network and ISP provider is sufficient 
to achieve Mirage's benefits in this scenario. Even when the attack throughput is $10$ Gbps, the median distance to the pushback links is only $4$ 
router hops and $2$ AS hops.

\section{Deployment}  %
\label{sec:deployment-comparison}

\begin{table*}[t]
\footnotesize
\centering
\begin{tabular}{|p{1.3in}|p{0.8in}|p{0.9in}|p{0.9in}|p{0.9in}|p{1.2in}|}
\hline {\bf \em Mechanism} & {\bf Mirage} & {\bf NetFence~\cite{liu:sigcomm10}} & {\bf TVA~\cite{yang:sigcomm05}} & {\bf Portcullis~\cite{parno:sigcomm07}} & {\bf Phalanx~\cite{dixon:nsdi08}} \\
\hline
\hline {\em Filtering} & {\bf IP dest based} & Capability based & Capability based & Capability based & Capability based \\
\hline {\em Router communication}  & {\bf Out of band}  & New header &  New header & New header  & New header \\
\hline {\em Fair queuing/\prateek{rate limiting}} & {\bf Yes} & {\bf Yes} & {\bf Yes} & {\bf Yes} & {\bf Yes} \\
\hline {\em Puzzle distribution} & Yes  & {\bf No}  &  {\bf No} (attack~\cite{parno:sigcomm07}) & Yes  &  Yes \\
\hline {\em Bandwidth-based fairness} & {\bf No}  & {\bf No}  &  {\bf No} & {\bf No}  &  Yes \\
\hline {\em Router Cryptography} & {\bf No} & Yes & Yes & Yes & {\bf No} \\
\hline {\em Other} & {\bf None} & Passport & Path identifiers & {\bf None} & Traffic forwarding via overlay network\\
\hline
\end{tabular}
\begin{security}
\vspace{-0.1in}
\end{security}
\caption{In-network support requirements for DDoS defense schemes.}
\begin{security}
\vspace{-0.1in}
\end{security}
\label{tbl:mechanism_comparison}
\end{table*}

\begin{table*}[t]
\footnotesize
\centering
\begin{tabular}{|p{1.5in}|p{0.9in}|p{0.8in}|p{0.8in}|p{0.9in}|p{1.15in}|}
\hline {\bf \em Deployment details} & {\bf Mirage} & {\bf NetFence~\cite{liu:sigcomm10}} & {\bf TVA~\cite{yang:sigcomm05}} &{\bf Portcullis~\cite{parno:sigcomm07}} & {\bf Phalanx~\cite{dixon:nsdi08}} \\
\hline
\hline {\em Source upgrades}  & {\bf Optional} & Required & Required & {Required} & {\bf Optional} \\
\hline {\em Destination  upgrades}  & {\bf Required} & {\bf Required} & {\bf Required} & {\bf Required} & {\bf Required} \\
\hline {\em Bottleneck router(s) \mbox{upgrades}} & {\bf Required} & {\bf Required} & {\bf Required} & {\bf Required} & {\bf Required} \\
\hline {\em Other router(s) upgrades} & {\bf None} & Yes  & Yes  & {\bf None} & {Yes (Overlay member)}  \\
\hline {\em Puzzle distribution} & {Akamai/DNS } & {\bf None} & {None (attack)} & \prateek{Akamai/DNS} & Akamai/DNS \\
\hline {\em Router upgrade type} & {\bf Configuration changes (may need more memory)} & \mbox{Software and} hardware changes \mbox{(by vendors)} & \mbox{Software and} hardware changes \mbox{(by vendors)}& \mbox{Software and} hardware changes \mbox{(by vendors)}& \mbox{Software and} hardware changes  \mbox{(by vendors)}  \\
\hline {\em Other} & {\bf None} & {\bf None} & {\bf None} & Trusted authority & Overlay network \\
\hline
\end{tabular}
\begin{security}
\vspace{-0.1in}
\end{security}
\caption{Affected deployment locations for DDoS defense schemes.}
\label{tbl:deployment_comparison}
\begin{security}
\vspace{-0.4in}
\end{security}
\end{table*}

Like %
previous DDoS mitigation schemes, our design
requires changes to certain components within the network. However, our work
differs from ``clean slate'' approaches which require large-scale changes to
the Internet infrastructure to achieve their benefits. In particular, our
experimental results show that only the victim server's local network and its
upstream ISP can deploy Mirage to defend against moderate scale attacks.
Moreover, Mirage is incentive compatible. Since the victim has a business
relationship with its upstream ISP, this may spur economic benefits for
adoption~\cite{goldberg:sigcomm11}. Incrementally deploying Mirage at ISPs that
are further upstream increases Mirage's resilience against attack.
Mirage %
requires the following modifications
to existing systems:  %

\paragraphb{ADNS:} Mirage requires changes to the server's
authoritative DNS server, to redirect clients to the puzzle server,
which can be hosted by \emph{existing} services such as Akamai or
Amazon S3.
Since the authoritative DNS server is typically owned and operated by
the service provider, many deployed systems (e.g., Akamai) leverage
the ADNS as an easy-to-modify location to instrument their designs.

\paragraphb{Server:}
Mirage can be deployed as a bump in the wire solution at the server; existing software
does not need to be modified. %
For example, Mirage software can %
simply bind the server's hopping IP addresses
to $0.0.0.0$.  Alternately, Mirage can also be deployed as a reverse proxy.

\paragraphb{Server's upstream network(s):} Mirage also requires changes to
networks. However, Mirage does not require widespread adoption in order for
servers to benefit. Changes to network infrastructure that is under direct
 influence of the service provider, such as its own network, and ISPs that
it directly pays for service, suffice to defend against moderate scale attacks.
In addition, these network changes do not %
rely on router vendors to incorporate new changes into their software, and can
instead be realized through configuration changes.
\prateek{However, depending on the point of deployment, the ACL mechanism may
need to be scaled (which can be done by the operator by installing more
memory, rather than needing to convince the vendor to change the router's
hardware/software design.}
These configuration changes
can be automated through the use of an IRSCP~\cite{irscp} to install ACLs within
the network. Mirage can alternatively leverage OpenFlow, with Mirage's functionality
implemented within the NOX controller. %
Several large ISPs have
already begun offering commercial services to allow customers to install prefixes
for blacklist filtering~\cite{attddosdefense}. %

\subsection{Comparison with other approaches}
Next, we contrast the deployment challenges of Mirage with %
previous proposals,
in terms of what changes need to be made to the infrastructure (network primitives), and
which players in the network need to instrument those changes (administrative boundaries).

\paragraphb{Adding new primitives to the network:} Deploying new functionality
in the network becomes easier when the changes to existing devices are small.
Protocols that require new primitives in routers require %
coordination with (and across) network vendors to be realized.
Table~\ref{tbl:mechanism_comparison} shows the type of changes required to network
devices to deploy several recently-proposed DDoS mitigation schemes.
First, NetFence, TVA, Portcullis, and Phalanx perform
filtering based on some form of a capability. This requires modifications to the
software of Internet routers to process and modify this capability. In contrast,
Mirage performs filtering based on the destination IP address, and communicates ACL entries to
routers using legacy protocols (iBGP feeds or configuration changes via sessions to an IRSCP).
Next, Mirage, Portcullis, and Phalanx require computational puzzles (proof of
work) in order to defend against attacks targeting the request channels. TVA does not use
computational puzzles, but this makes it vulnerable to the Denial of Capability (DoC) attack~\cite{parno:sigcomm07}. %
NetFence requires defenses to prevent the IP spoofing attack (such as Passport) to be
widely deployed in order to secure the request channel.
\dongho{In addition to computation-based fairness using computational puzzles,
Phalanx also implicitly adopts bandwidth-based fairness~\cite{walfish:sigcomm06},
an approach which is known to induce congestion collapse in the network.}
Finally, the non-Mirage mechanisms require additional functionality
from the network, such as 1) cryptography at the routers for NetFence, TVA, and
Portcullis, 2) insertion of path identifiers for TVA, and 3) overlay network to 
forward traffic for Phalanx.

\paragraphb{Deploying changes across administrative boundaries:} Deploying new
functionality in the network becomes easier when fewer participants have to
cooperate in deploying those changes.  Protocols that require clients to
install new software, or require widespread upgrades to routers can complicate
deployment.  Table~\ref{tbl:deployment_comparison} describes the location where
changes would need to be realized within the network.  First, only Mirage and
Phalanx do not require source hosts to upgrade.  Second, all mechanisms require
destinations to upgrade, and also require deployment at the bottleneck routers
(points of congestion).  However, TVA and NetFence also require other routers to
upgrade, for inserting path identifiers and for defending against IP spoofing;
this means that it is not possible for an ISP to unilaterally deploy NetFence.
Similarly, Phalanx also requires either an external overlay network to absorb the
DDoS attack, or requires additional routers to upgrade and participate in the
overlay scheme. \prateek{Portcullis requires all Internet routers to trust a single entity.}
Finally, Mirage, Portcullis, and Phalanx require an additional
deployment component to distribute puzzles. %
However, we note that puzzle distribution in Mirage,
Portcullis and Phalanx does not require global cooperation from ISPs.
A Victim end host can easily use
existing CDNs like Akamai, which can be given financial incentives to
participate. Finally, we note that Mirage is the {\em only} scheme in which
functionality required from the routers is already deployed in today's routers\prateek{,
though this functionality (ACLs) may need to be scaled depending on the point
of deployment}. Routers already have support for ACLs and fair queuing, which
can be enabled with relatively straightforward configuration
commands. %
Scaling memory requirements for ACLs can be done by the operator
by installing more memory at the router, rather than requiring
router vendor's cooperation.

\section{Discussion}
\label{sec:discussion}
Here, we discuss some ramifications of our design.

\subsection{Applicability to non-web applications} \prateeknew{Mirage is ideally suited to defend 
against web applications, since its JavaScript mechanism obviates the need for client 
software.  However, Mirage can also be used to defend against non-web applications, 
at the cost of upgrading clients. The JavaScript mechanism can be replaced by 
client software that provides equivalent functionality of fetching/solving puzzles. 
We note that even in this setting, Mirage provides stronger deployment properties 
compared with previous work, since it provides security benefits when only the victim's 
upstream ISP cooperates (as illustrated in Table~\ref{tbl:deployment_comparison})}.  

\begin{figure}
\centering
\includegraphics[width=0.35\textwidth]{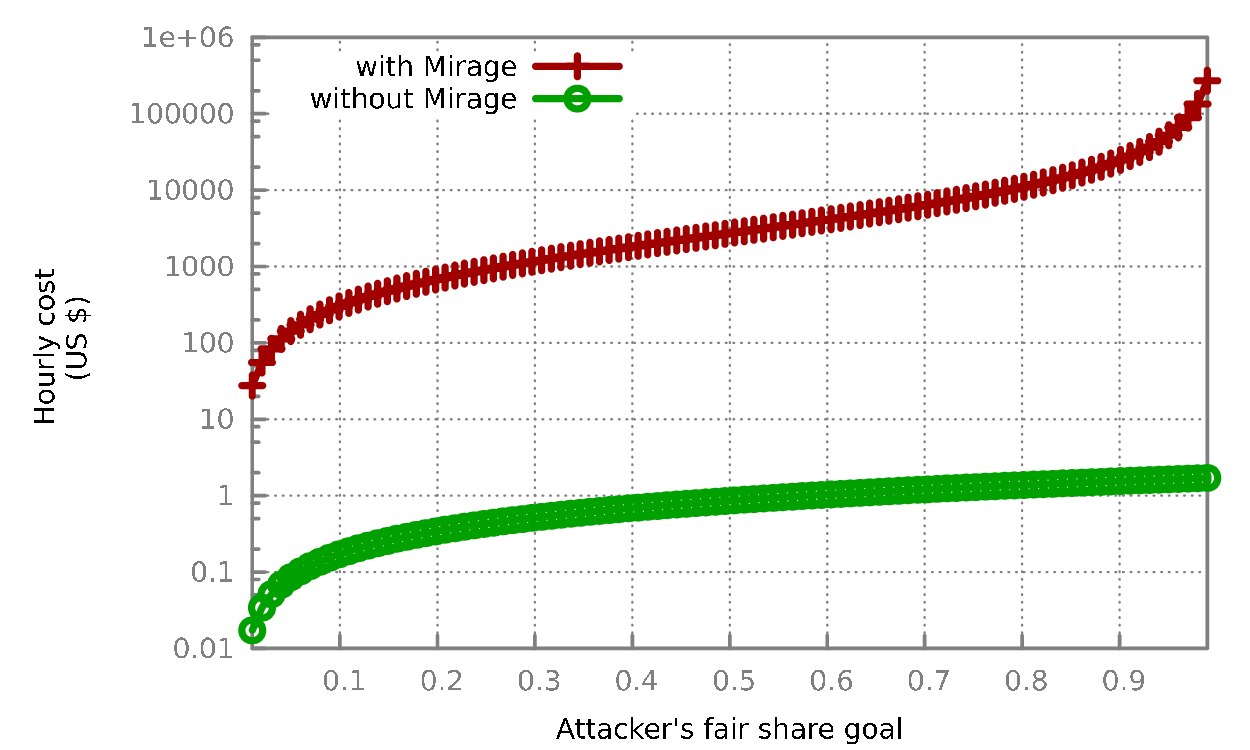}
\vspace{-0.1in}
\caption{Hourly DDoS attack cost as a function of desired share of victim's capacity. Mirage raises the attack cost 
by several orders of magnitude.}
\label{fig:cost-analysis}
\vspace{-0.2in}
\end{figure}

\subsection{Cost analysis using EC2 pricing} \prateeknew{ Next, we perform a cost analysis 
of launching a DDoS attack by considering (a) Amazon EC2 prices for computation and data 
transfer~\cite{ec2-pricing}, and (b) a data center environment which serves hundreds 
of thousands of users in short time periods~\cite{devoflow} as the victim server. Let us suppose 
that the victim has a $1$~Gbps link to its service provider, and that the computational capability 
of each honest user is 1 EC2 compute unit (equivalent to 1.0-1.2 GHz 2007 Xeon processor). 
Without Mirage, the cost of obtaining a fraction $x$ of the victim's network capacity is 
proportional only to the required EC2 data transfer. With Mirage, the attacker is forced to spend 
extra resources for solving computational puzzles. Figure~\ref{fig:cost-analysis} depicts the 
hourly cost of launching a DDoS attack in US dollars, as a function of the attacker's desired 
share of the victim's network capacity. We can see that Mirage raises the attack cost by several 
orders of magnitude. In future work, we will extend this analysis by considering botnet rent 
prices for computation and bandwidth. 
}

\subsection{Adaptive network defense} To improve performance under small and
moderate sized attacks, we describe an adaptive mechanism that utilizes network
support in proportion to the strength of the attack.  In
particular, we organize the active set of IP addresses as a binary tree, with
nodes in the tree representing IP addresses, and the number of  leaves
equaling the number of potential clients.  Levels in the tree represent
priorities: IP addresses at level $i$ are assigned half the priority as
compared to IP addresses at level $i+1$.  We modify our puzzle server to return
multiple IP addresses, one from each level in the tree.  Source end
hosts first try the IP address  corresponding to the root of the tree (level
$0$), and if they are unable to get good service, they adaptively switch to
the next level in the binary tree. This has the advantage that  the router only
needs to maintain {\em per aggregate state} in the access control lists
while performing fair queuing.
Similarly, the ACL entries could also be added adaptively by the
destination end host - it  starts off by adding the root of the tree, and then
progressively adding next  level IP addresses as needed. %
This reduces the ACL list size.
\subsection{Authorized users and classful services} We can extend our
architecture to enable destination end hosts to provide higher priority to authorized
users. Firstly, victim servers could communicate private
IP addresses to authorized users using out-of-band communications. Secondly,
authorized users could use Mirage to authenticate to the victim server and
receive private IP addresses, or \emph{hints} to solve computational puzzles.
Finally, authorized users could authenticate themselves to the puzzle servers,
using a server-supplied cookie (possibly cached from last visit by the browser).
The authentication could be performed in zero knowledge, with
no information about client credentials being revealed to the puzzle server. Upon
authentication, the puzzle server could provide either private IP addresses or
hints for puzzle to the authorized users.

We note that our architecture can enable a useful scenario
where a sensitive server's IP address is visible only to authorized users.
Here, the server would need to supply the cookie to authorized users via
an out of band channel, and the puzzle server would respond only after
authentication. The  server's IP address is practically invisible to the rest
of the Internet, and is thus largely immune to threats like bot infections due
to random IP scanning.
 \subsection{Insider collusion attacks}
 A potential concern in Mirage  is  the collusion between a compromised end host in the victim network and the attacker. Suppose that the victim network
 has two nodes. The compromised end host could install an ACL at the upstream AS with $k$ entries, with the attackers having full knowledge of
 the %
active IP addresses for the compromised %
host. The attackers could then send traffic to the $k$ IP addresses listed in the ACL.
 Due to per destination fair queueing at the upstream routers, the attackers would gain an overwhelming share of the traffic at the victim link and DoS the  honest end host.
In practice, the operator can deal with this through appropriate network management.
One %
management policy to solve the problem could be to bound the size of the ACL per victim.
%
 %
%

%
%
%
%
%
%
%
%
%
%
%
%
%
%
%
%
%

%
%
%
%
%
%

%
%
%

%
\subsection{Prefix/path/location hopping}
While our design focuses on hopping of IP addresses, it may be possible
to apply this technique to other components of network protocols.
For example, it may be possible to perform {\em prefix hopping}, by
having router prefixes changing with time,
{\em path hopping}, by changing the set of Internet paths used to route
to a destination,
and {\em location hopping}, by leveraging virtual machine migration techniques
to relocate services on-demand in the presence of a DoS attack.
These techniques may assist in providing improved security properties in present
and future Internet architectures.

\section{Conclusion}
\label{sec:conclusion}

Mirage uses an analog of {\em frequency hopping} from wireless networks:
hosts vary their IP address through a pseudorandom sequence to evade
attacks by unauthorized hosts.
From theoretical analysis, simulations, and experiments from a prototype
implementation, we find that Mirage achieves comparable performance to previous
DDoS mitigation mechanisms.  %
However, Mirage has improved deployment properties over previous designs. 
Mirage does not require an external overlay network, or
a trusted Internet authority, or defenses against IP spoofing.
In particular, Mirage is incrementally deployable; a victim server and its upstream ISP can deploy Mirage
and defend against moderate scale attacks, without  %
requiring source end hosts
to install any software.  %
\section*{Acknowledgments}
We are grateful to Prateek Saxena and Devdatta Akhawe for discussions 
about our JavaScript mechanism. Robin Sommer, Vern Paxson, Brian Tierney, 
and Chin Guok provided us with helpful feedback. %

\balance
{\bibliographystyle{abbrv}
\bibliography{refs-short}
}

\end{document}